%% file: JpsiX_psi3686X_BESIII_28Jan2025.tex
\documentclass[reprint,amsmath,amssymb,aps,prd]{revtex4-1}
\usepackage{graphicx}
\usepackage{dcolumn}
\usepackage{bm}
\usepackage{rotating}
\usepackage{xcolor}
\usepackage[unicode]{hyperref} 

\begin{document}
\normalsize
\parskip=5pt plus 1pt minus 1pt

\title{\boldmath Measurement of the inclusive cross sections of prompt $J/\psi$ and $\psi(3686)$ production in $e^{+}e^{-}$ annihilation from $\sqrt{s}=3.808$ to $4.951$~GeV}
\author{\input{authorlist_2023-05-10}}
\date{January 28, 2025}

\begin{abstract}
  The inclusive cross sections of prompt $J/\psi$ and $\psi(3686)$ production are measured at center-of-mass energies from 3.808 to 4.951~GeV.  The dataset used is 22~fb$^{-1}$ of $e^{+}e^{-}$ annihilation data collected with the BESIII detector operating at the BEPCII storage ring. The results obtained are in agreement with the previous BESIII measurements of exclusive $J/\psi$ and $\psi(3686)$ production. The average values obtained for the cross sections measured in the center-of-mass energy ranges from 4.527 to 4.951~GeV for $J/\psi$ and from 4.843 to 4.951~GeV for $\psi(3686)$, where the impact of known resonances is negligible, are $14.0\pm1.7\pm3.1$~pb and $15.3\pm3.0$~pb, respectively. For $J/\psi$, the first and the second uncertainties are statistical and systematic, respectively. For $\psi(3686)$, the uncertainty is total. These values are useful for testing charmonium production models.
\end{abstract}

\maketitle

\section{INTRODUCTION}

Despite decades of dedicated studies, charmonium production in different processes and over a wide range of energies remains a challenge for our understanding of quantum chromodynamics (QCD). A vast amount of data on charmonium production has already been collected in $e^+e^-$ annihilation experiments, as well as in photoproduction and hadroproduction. Interest in the mechanisms of charmonium production was boosted by the discovery of a number of exotic hadron states beyond the conventional quark model because charmonia are present among their decay products. The mechanisms of quarkonium production are also of special interest in the physics programs of future projects at both low~\cite{PANDA,AMBER,SPD,EIC} and high energies~\cite{FCC,ILC,CEPC}.

Non-relativistic QCD (NRQCD)~\cite{Bodwin:1994jh} provides the most rigorous approach to describe charmonium production. In NRQCD, the production cross section factorizes into perturbative short-distance coefficients and non-perturbative long-distance matrix elements (LDMEs). The short-distance coefficients describe the production of a $c\bar{c}$ pair in a particular color and angular-momentum state, while the LDMEs define the probability for this pair to evolve into a particular charmonium state. The LDMEs extracted from the different sets of experimental data~\cite{Gong:2013JpsiPsi,Shao:2015JpsiPsi,Baranov:2019ch,Brambilla:2022qu,Butenschoen:2023psi3686} should be universal, but the validity of this claim remains unclear in view of the large uncertainties. For example, B-factory data on prompt inclusive $J/\psi$ production at $\sqrt{s} = 10.6$~GeV~\cite{Aubert:2001pd,Pakhlov:2009nj,Briere:2004ug} were described successfully with next-to-leading order (NLO) NRQCD perturbative calculations. However, extrapolation to lower energies~\cite{Gong:2019rpd} give results much smaller than the cross sections of the exclusive $e^+e^- \rightarrow \pi^+\pi^-J/\psi$ process measured at center-of-mass energies from 3.770 to 4.600~GeV by BESIII~\cite{Ablikim:2016qzw}. This inconsistency is most likely caused by the inapplicability of perturbative calculations close to the $J/\psi$ production threshold. New experimental results on inclusive charmonium production in $e^+e^-$ annihilation at low energies will certainly be useful for understanding the limits of applicability of the NRQCD approach. The behavior of the inclusive cross sections in the region of numerous exotic charmonium-like states will also provide information about the properties of these resonances.

In this paper, the inclusive cross sections of prompt $J/\psi$ and $\psi(3686)$ production are measured at center-of-mass energies between 3.808 and 4.951~GeV~\cite{BeamEnergy2011_2014,BeamEnergy2017_2019,BeamEnergyLuminosity2020_2021} using $e^{+}e^{-}$ annihilation data corresponding to an integrated luminosity of 22~fb$^{-1}$~\cite{BeamEnergyLuminosity2020_2021,Luminosity2011_2017}. The analyzed data was collected between 2011 and 2021 with the BESIII detector operating at the BEPCII storage ring. Only $J/\psi$ and $\psi(3686)$ mesons produced directly in $e^{+}e^{-}$ annihilation at the nominal energy are treated as signal. Therefore mesons originating from decays of conventional charmonia as well as those produced via initial state radiation (ISR), both from ISR return to the $J/\psi$ and $\psi(3686)$ resonances as well as ISR to the lower-energy continuum, are excluded from the analysis.

\section{DETECTOR AND MONTE CARLO SIMULATION}

The BESIII detector~\cite{Ablikim:2009aa} records symmetric $e^+e^-$ collisions 
provided by the BEPCII storage ring~\cite{Yu:IPAC2016-TUYA01} in the center-of-mass energy range from 2.000 to 4.951~GeV,
with a peak luminosity of $1 \times 10^{33}\;\text{cm}^{-2}\text{s}^{-1}$ 
achieved at $\sqrt{s} = 3.770\;\text{GeV}$. 
BESIII has collected large data samples in this energy region~\cite{Ablikim:2019hff}.
The cylindrical core of the \mbox{BESIII} detector covers 93\% of the full solid angle and consists of a helium-based
 multilayer drift chamber~(MDC), a plastic scintillator time-of-flight
system~(TOF), and a CsI(Tl) electromagnetic calorimeter~(EMC),
which are all enclosed in a superconducting solenoidal magnet
providing a 1.0~T magnetic field. The solenoid is supported by an
octagonal flux-return yoke with resistive plate counter muon
identification modules interleaved with steel. 
The charged-particle momentum resolution at $1~{\rm GeV}/c$ is
$0.5\%$, and the $dE/dx$ resolution is $6\%$ for electrons
from Bhabha scattering. The EMC measures photon energies with a
resolution of $2.5\%$ ($5\%$) at $1$~GeV in the barrel (end cap)
region. The time resolution in the TOF barrel region is 68~ps, while
that in the end cap region is 110~ps. The end cap TOF
system was upgraded in 2015 using multi-gap resistive plate chamber
technology, providing a time resolution of 60~ps~\cite{etof}.  
About 75\% of the data used here benefits from this upgrade.

Simulated samples produced with a {\sc
geant4}-based~\cite{geant4} Monte Carlo (MC) package, which
includes the geometric description~\cite{detvis} of the BESIII detector and the
detector response, are used to determine reconstruction efficiencies
and to estimate backgrounds. The simulation models the beam
energy spread and ISR in the $e^+e^-$
annihilations with the generator {\sc
kkmc}~\cite{ref:kkmc}.
The inclusive MC samples include the ISR production of vector charmonium(-like) states,
and the continuum processes incorporated in {\sc
kkmc}.
All particle decays are modelled with {\sc
evtgen}~\cite{ref:evtgen} using branching fractions 
either taken from the
Particle Data Group (PDG)~\cite{pdg}, when available,
or otherwise estimated with {\sc lundcharm}~\cite{ref:lundcharm}.
Final state radiation~(FSR)
from charged final state particles is incorporated using the {\sc
  photos} package~\cite{photos}.

\section{METHODOLOGY}

The total number of events with $J/\psi$ and $\psi(3686)$ in the final state can be determined accurately from the reconstruction of the decays $J/\psi \rightarrow \mu^{+}\mu^{-}$ and $\psi(3686)  \rightarrow \pi^{+}\pi^{-} J/\psi, J/\psi \rightarrow l^{+}l^{-}$~($l = e, \mu$), respectively. For the process $e^{+}e^{-}\rightarrow J/\psi_{\textmd{prompt}}X$, the reconstruction of the $J/\psi$ meson via a pair of electrons is not used because it does not improve the result significantly, due to large Bhabha backgrounds.  
Since only $J/\psi$ ($\psi(3686)$) mesons produced directly in $e^{+}e^{-}$ annihilation at the nominal energy are considered as the signal, the following cases, indicated below, are excluded from our analysis.

First, $J/\psi$ mesons are produced in the decays of the conventional charmonium states $\psi(3686)$ and $\chi_{cJ}$~(\textit{J} = 1, 2). The number of such events is estimated from the reconstruction of the well-known decays $\psi(3686)  \rightarrow \pi^{+}\pi^{-} J/\psi$ and $\chi_{cJ} \rightarrow \gamma J/\psi$, taking into account reconstruction efficiency for each decay channel. 
We also consider other decay channels of the $\psi(3686)$ which have a $J/\psi$ in the final state, but exclude the decays $\psi(3686)  \rightarrow \gamma \chi_{cJ}, \chi_{cJ} \rightarrow \gamma J/\psi$~(\textit{J} = 1, 2) to avoid double counting. Other established charmonium states such as $\psi(3770)$, $\chi_{c0}$, etc., are ignored due to their negligible contribution. Thus, the inclusive cross sections of $J/\psi$ production around the $\psi(3770)$ peak is 2 orders of magnitude less than around the $\psi(3686)$ peak~\cite{JpsiX_3770}. While the radiative decays of the $\chi_{c0}$ meson in the $J/\psi$ meson are suppressed by an order of magnitude compared to similar decays of the $\chi_{c1}$ and $\chi_{c2}$ mesons~\cite{pdg}.  However, the $J/\psi$ mesons produced in the decay of exotic states like $\psi(4230)$, $Z_c(3900)$, etc., are treated as signal in this analysis since information about the properties of these resonances is also of interest. 

Second, $J/\psi$ mesons are produced via the ISR process returning to the $J/\psi$ resonance. Most of such events are rejected by the event selection criteria described below. The residual percentage of such events is estimated and subtracted based on the reconstruction of the reaction $e^{+}e^{-}\rightarrow \gamma_{\textmd{ISR}}J/\psi$, where $\gamma_{\textmd{ISR}}$ denotes one or more ISR photons, taking into account the reconstruction efficiency obtained from the MC simulation.

Third, $J/\psi$ ($\psi(3686)$) mesons are generated through the ISR process returning to the $\psi(3686)$ resonance. The contribution of this mechanism to the inclusive production of the $J/\psi$ meson is estimated and subtracted basing on the reconstruction of the reaction $e^{+}e^{-}\rightarrow \gamma_{\textmd{ISR}}\psi(3686)$   and calculation of reconstruction efficiency using MC simulation. For the inclusive production of the $\psi(3686)$ meson, most of these events are rejected by the event selection criteria described below. The residual percentage of such events is estimated and subtracted as the $e^{+}e^{-}\rightarrow \gamma_{\textmd{ISR}}\psi(3686)$ background from the observed number of $e^{+}e^{-}\rightarrow \psi(3686)X$ events, where $X$ denotes the usual inclusive final state. In the case of the inclusive $J/\psi$ meson production, this background is also subtracted to avoid double counting.

Thus, for each energy point, the observed cross section of the process $e^{+}e^{-}\rightarrow J/\psi_{\text{prompt}}X$, which is denoted by $\sigma^{O}_{J/\psi}$, is expressed as: 
\begin{align} 
  \sigma^{O}_{J/\psi} = \frac{1}{\mathcal{L}} \left[ Y_{J/\psi X} - Y_{\psi X} - Y_{\gamma_{\textmd{ISR}}\psi} - \sum_{J=1,2} Y_{\chi_{cJ} X} \right],  
  \label{CS_JpsiX_prompt}
\end{align}
where $\mathcal{L}$ represents the integrated luminosity, $Y_{J/\psi X}$ is the inclusive $J/\psi$ production yield, $Y_{\psi X}$ is the inclusive $\psi(3686)$ production yield with $\psi(3686) \rightarrow J/\psi X$, $Y_{\gamma_{\textmd{ISR}}\psi}$ is the yield from ISR return to the $\psi(3686)$ resonance with $\psi(3686) \rightarrow J/\psi X$, and $Y_{\chi_{cJ} X}$~(\textit{J} = 1, 2) is the inclusive $\chi_{cJ}$ production yield with $\chi_{cJ} \rightarrow \gamma J/\psi$. 

The inclusive $J/\psi$ production yield is given by the following formula:
\begin{equation}
Y_{J/\psi X} = \frac{N_{J/\psi X} - \mathcal{R}_{\gamma_{\textmd{ISR}}J/\psi} \, N_{\gamma_{\textmd{ISR}}J/\psi }}{\epsilon_{J/\psi X} \, \mathcal{B}_{J/\psi \rightarrow \mu^{+}\mu^{-}}},
   \label{CS_JpsiX}
 \end{equation}
 where $N_{J/\psi X}$ is the observed number of  $e^{+}e^{-}\rightarrow J/\psi X$ events; $N_{\gamma_{\textmd{ISR}}J/\psi }$ is the observed number of $e^{+}e^{-}\rightarrow \gamma_{\textmd{ISR}}J/\psi$ events; $\mathcal{R}_{\gamma_{\textmd{ISR}}J/\psi}$, calculated from {\sc kkmc} samples of the process $e^{+}e^{-}\rightarrow \gamma_{\textmd{ISR}}J/\psi$, is the ratio of the number of events that satisfy the selection criteria for processes $e^{+}e^{-}\rightarrow J/\psi X$ and $e^{+}e^{-}\rightarrow \gamma_{\textmd{ISR}}J/\psi$; $\epsilon_{J/\psi X}$ is the reconstruction efficiency of the $J/\psi$ meson in $e^{+}e^{-}\rightarrow J/\psi X$ events, and $\mathcal{B}_{J/\psi \rightarrow \mu^{+}\mu^{-}}$ is the branching fraction of the decay $J/\psi \rightarrow \mu^{+}\mu^{-}$ taken from the PDG~\cite{pdg}.

A uniform designation is introduced for the quantities used to measure the cross section of both processes  $e^{+}e^{-}\rightarrow J/\psi_{\textmd{prompt}}X$ and $e^{+}e^{-}\rightarrow \psi(3686)_{\textmd{prompt}}X$.  The labels~``$\mu$",  ``$e$", and~``$l$" indicate that the corresponding value is obtained for decays of the $J/\psi$ meson into a pair of muons, electrons, and both leptons, respectively. Thus, the inclusive $\psi(3686)$ production yield $Y_{\psi X}$ with $\psi(3686) \rightarrow J/\psi X$ is given by the formula:
 \begin{equation}
Y_{\psi X} = \frac{\left(N^{\mu}_{\psi X} - \mathcal{R}^{\mu}_{\gamma_{\textmd{ISR}}\psi} \, N^{\mu}_{\gamma_{\textmd{ISR}}\psi }\right) \mathcal{\tilde B}_{\psi  \rightarrow J/\psi X}}{\epsilon^{\mu}_{\psi X} \, \mathcal{B}_{\psi  \rightarrow \pi^{+}\pi^{-} J/\psi} \, \mathcal{B}_{J/\psi \rightarrow \mu^{+}\mu^{-}}}, 
   \label{CS_psiToJpsi}
 \end{equation}
where $N^{\mu}_{\psi X}$ is the observed number of $e^{+}e^{-}\rightarrow  \psi(3686) X$ events; $N^{\mu}_{\gamma_{\textmd{ISR}}\psi }$ is the observed number of $e^{+}e^{-}\rightarrow \gamma_{\textmd{ISR}}\psi(3686)$ events; $\mathcal{R}^{\mu}_{\gamma_{\textmd{ISR}}\psi}$, which is calculated using {\sc kkmc} samples of the process $e^{+}e^{-}\rightarrow \gamma_{\textmd{ISR}}\psi(3686)$, is the ratio of the number of events that satisfy the selection criteria for processes $e^{+}e^{-}\rightarrow \psi(3686) X$ and $e^{+}e^{-}\rightarrow \gamma_{\textmd{ISR}}\psi(3686)$; $\epsilon^{\mu}_{\psi X}$ is the reconstruction efficiency of the $\psi(3686)$ meson in $e^{+}e^{-}\rightarrow \psi(3686) X$ events; $\mathcal{B}_{\psi  \rightarrow \pi^{+}\pi^{-} J/\psi} $ is the branching fraction of the $\psi(3686)  \rightarrow \pi^{+}\pi^{-} J/\psi$  decay, and $ \mathcal{\tilde B}_{\psi  \rightarrow J/\psi X}$  is the branching fraction of the $\psi(3686)  \rightarrow J/\psi X$ decay excluding the decays $\psi(3686)  \rightarrow \gamma \chi_{cJ} \rightarrow \gamma(\gamma J/\psi)$:
 \begin{equation}
   \mathcal{\tilde B}_{\psi  \rightarrow J/\psi X} = \mathcal{B}_{\psi  \rightarrow J/\psi X} - \sum_{J=1,2} \mathcal{B}_{\psi  \rightarrow \gamma \chi_{cJ}} \, \mathcal{B}_{\chi_{cJ} \rightarrow \gamma J/\psi}.
   \label{BF_psiToJpsiX}
 \end{equation}
 
The yield from ISR return to the $\psi(3686)$ resonance $Y_{\gamma_{\textmd{ISR}}\psi}$ with $\psi(3686) \rightarrow J/\psi X$ is given by: 
  \begin{equation}
   Y_{\gamma_{\textmd{ISR}}\psi} = \frac{N^{\mu}_{\gamma_{\textmd{ISR}} \psi} \, \mathcal{\tilde B}_{\psi  \rightarrow J/\psi X}}{\epsilon_{\gamma_{\textmd{ISR}} \psi} \, \mathcal{B}_{\psi  \rightarrow \pi^{+}\pi^{-} J/\psi} \, \mathcal{B}_{J/\psi \rightarrow \mu^{+}\mu^{-}}},
   \label{CS_psiISR}
 \end{equation}
where $\epsilon_{\gamma_{\textmd{ISR}} \psi}$ is the reconstruction efficiency of the $\psi(3686)$ meson in $e^{+}e^{-}\rightarrow \gamma_{\textmd{ISR}} \psi(3686)$ events. 

The inclusive $\chi_{cJ}$~(\textit{J} = 1, 2) production yield  $Y_{\chi_{cJ} X}$ with $\chi_{cJ} \rightarrow \gamma J/\psi$ is given by: 
 \begin{equation}
   Y_{\chi_{cJ} X} = \frac{N_{\chi_{cJ} X}}{\epsilon_{\chi_{cJ} X} \, \mathcal{B}_{J/\psi \rightarrow \mu^{+}\mu^{-}}}, 
   \label{CS_chiToJpsi}
 \end{equation}
where $N_{\chi_{cJ} X}$ is the observed number of $e^{+}e^{-}\rightarrow  \chi_{cJ} X$ events, and $\epsilon_{\chi_{cJ} X}$ is the reconstruction efficiency of the $\chi_{cJ}$ meson in $e^{+}e^{-}\rightarrow  \chi_{cJ} X$ events.

Similarly, the observed cross section of the process $e^{+}e^{-}\rightarrow \psi(3686)_{\textmd{prompt}}X$, which is denoted by $\sigma^{O}_{\psi, l}$,  for each energy point is expressed as: 
 \begin{align}
  \sigma^{O}_{\psi, l} = \frac{N^{l}_{\psi X} - \mathcal{R}^{l}_{\gamma_{\textmd{ISR}}\psi} \, N^{l}_{\gamma_{\textmd{ISR}}\psi }}{\mathcal{L} \, \epsilon^{l}_{\psi X} \, \mathcal{B}_{\psi  \rightarrow \pi^{+}\pi^{-} J/\psi} \, \mathcal{B}_{J/\psi \rightarrow l^{+}l^{-}}},  
  \label{CS_psiX_prompt}
 \end{align}
where $l = e, \mu$ denotes the lepton decay mode of  the $J/\psi$ meson. Next, the observed cross sections obtained using the di-muon ($\sigma^{O}_{\psi, \mu}$) and di-electron ($\sigma^{O}_{\psi, e}$) decay modes are averaged only taking into account independent uncertainties. The average observed cross section is further denoted by $\sigma^{O}_{\psi}$. The uncertainties are clearly divided into independent and common. The independent uncertainty for a specific lepton decay mode is obtained by quadratically summing statistical uncertainty and independent contributions to systematic uncertainty. The common uncertainty, which is further calculated only for the Born cross section, is the quadratic sum of the common contributions to systematic uncertainty.

The Born cross section, $\sigma^{B}(s)$, is obtained from the observed cross section by applying radiative corrections:
\begin{align}
  \sigma^{B}(s) = \frac{\sigma^{O}(s)}{[1 + \delta(s)] \, |1-\Pi(s)|^{-2}}= \frac{\sigma^{D}(s)}{|1-\Pi(s)|^{-2}},
    \label{CS_Born}
\end{align}
where $1 + \delta(s)$ denotes the ISR correction factor, taking into account ISR return to the continuum, $|1-\Pi(s)|^{-2}$ is the vacuum polarization factor taken from a QED calculation \cite{ISR_formula}, and $\sigma^{D}(s)$ signifies the total dressed cross section without the correction for vacuum polarization effects. 

To obtain the ISR correction factor $1 + \delta(s)$ at each energy point we apply the following iteration procedure:
\begin{align}
  \sigma^{D}_{i+1}(s) = \frac{\sigma^{O}(s)}{[1 + \delta(s)]_{i}},
  \label{Iter_def}
\end{align}
\begin{align}
 [1 + \delta(s)]_{i} = \frac{\int \sigma_{i}^{D}(s-sx) \, F(x,s) \, dx}{\sigma_{i}^{D}(s)},
  \label{Delta_def}
\end{align}
where $i$ is the iteration number, $s$ is the square of the center-of-mass energy, $x$ is the fraction of energy carried by the ISR photons, and $F(x,s)$ is the ISR function \cite{ISR_formula}. The function $F(x,s)$  is given by the following formula:

 \begin{multline} 
   F(x,s) = x^{t-1} \, t(1+\Delta) + x^{t}\left(-t-\frac{t^2}{4}\right) \\
   + x^{t+1}\left(\frac{t}{2}-\frac{3}{8}t^2\right) + O(x^{t+2} \, t^2), 
   \label{F_ISR}
 \end{multline}
 where $\Delta$ is defined as
 \begin{equation}
   \Delta = \frac{\alpha}{\pi}\left(\frac{\pi^2}{3}-\frac{1}{2}\right) + \frac{3}{4}t + t^2\left(\frac{9}{32}-\frac{\pi^2}{12}\right),
 \label{F_delta}
\end{equation}
and $t$ is calculated using the fine-structure constant, $\alpha$,
and the electron mass, $m_{e}$, as:
\begin{equation}
  t = \frac{2\alpha}{\pi}\left(\ln\frac{s}{m_{e}^2}-1\right).
  \label{F_t}
\end{equation}
The integration is carried out numerically in the range down to the $J/\psi$ peak with a relative accuracy better than $10^{-6}$. The cross section line-shape, $\sigma_{i}^{D}(s)$, for the iteration procedure is obtained from the fitting of an appropriate function to the measured dressed cross section. The fitting function is used to provide a reasonably smooth interpolation of the experimental points when calculating the integral.

\section{EVENT SELECTION}

In the analysis, 49 data samples with center-of-mass energies from 3.808 to 4.951~GeV are used. The center-of-mass energies~\cite{BeamEnergy2011_2014,BeamEnergy2017_2019,BeamEnergyLuminosity2020_2021} and corresponding integrated luminosities~\cite{BeamEnergyLuminosity2020_2021,Luminosity2011_2017} are given in Table~\ref{CS}. The event selection procedure described below is applied to each data sample. More details for the two samples with high integrated luminosity, at $\sqrt{s}=4.226$ and $4.682$~GeV, are provided to illustrate the event selection procedure.

\begin{table*}[htbp]
\caption{The center-of-mass energies $\sqrt{s}$, integrated luminosities $\mathcal{L}$, vacuum polarization correction $|1-\Pi|^{-2}$, ISR correction $(1 + \delta)$, and observed $\sigma^{O}$ and Born $\sigma^{B}$ cross sections of the processes $e^{+}e^{-}\rightarrow J/\psi_{\textmd{prompt}} X$ and $e^{+}e^{-}\rightarrow \psi(3686)_{\textmd{prompt}} X$ denoted by ``$ J/\psi$" and ``$\psi$",  respectively. For the $\sigma^{O}_{J/\psi}$ and $\sigma^{B}_{J/\psi}$, the first uncertainties are statistical, and the second are systematic. For the $\sigma^{O}_{\psi}$, the uncertainties are independent. For the $\sigma^{B}_{\psi}$,  the first uncertainties are independent, and the second are common.}
\begin{center}
\begin{tabular}{ccccccccc}\hline\hline
$\sqrt{s}$~(GeV) & $\mathcal{L}$~(pb$^{-1}$) & $|1-\Pi|^{-2}$ & $\sigma^{O}_{J/\psi}$~(pb) & $(1 + \delta)_{J/\psi}$ & $\sigma^{B}_{J/\psi}$~(pb) & $\sigma^{O}_{\psi}$~(pb) & $(1 + \delta)_{\psi}$ & $\sigma^{B}_{\psi}$~(pb) \\\hline
$3.808$ & $50.54$ & $1.056$ & $\textcolor{white}{.00}37\pm34\textcolor{white}{.}\pm33\textcolor{white}{.}$ & $0.929$ & $\textcolor{white}{.00}38\pm34\textcolor{white}{.}\pm34\textcolor{white}{.}$ & $-$ & $-$ & $-$ \\\hline
$3.896$ & $52.61$ & $1.049$ & $\textcolor{white}{.00}14\pm27\textcolor{white}{.}\pm18\textcolor{white}{.}$ & $0.930$ & $\textcolor{white}{.00}14\pm28\textcolor{white}{.}\pm18\textcolor{white}{.}$ & $-$ & $-$ & $-$ \\\hline
$4.008$ & $482.0$ & $1.044$ & $\textcolor{white}{.00}41\pm8.0\pm12\textcolor{white}{.}$ & $0.921$ & $\textcolor{white}{.00}43\pm9.0\pm13\textcolor{white}{.}$ & $\textcolor{white}{00}1.0\pm4.2$ & $0.869$ & $\textcolor{white}{00}1.2\pm4.6\pm0.0$ \\\hline
$4.085$ & $52.86$ & $1.052$ & $\textcolor{white}{.000}6\pm23\textcolor{white}{.}\pm9.0$ & $0.900$ & $\textcolor{white}{.000}7\pm24\textcolor{white}{.}\pm9.0$ & $\textcolor{white}{00}2.0\pm8.1$ & $0.871$ & $\textcolor{white}{00}2.2\pm8.8\pm0.0$ \\\hline
$4.129$ & $401.5$ & $1.052$ & $\textcolor{white}{0}24.0\pm8.0\pm9.0$ & $0.874$ & $\textcolor{white}{0}26.0\pm9.0\pm9.0$ & $\textcolor{white}{0}10.4\pm4.1$ & $0.868$ & $\textcolor{white}{0}11.4\pm4.4\pm0.3$ \\\hline
$4.158$ & $408.7$ & $1.053$ & $\textcolor{white}{00}5.0\pm8.0\pm8.0$ & $0.848$ & $\textcolor{white}{00}6.0\pm9.0\pm9.0$ & $\textcolor{white}{00}8.4\pm3.9$ & $0.864$ & $\textcolor{white}{00}9.2\pm4.3\pm0.3$ \\\hline 
$4.178$ & $3189.0$ & $1.054$ & $\textcolor{white}{0}34.5\pm2.9\pm8.3$ & $0.820$ & $\textcolor{white}{0}39.9\pm3.3\pm9.7$ & $\textcolor{white}{0}15.0\pm2.5$ & $0.859$ & $\textcolor{white}{0}16.6\pm2.7\pm0.6$ \\\hline
$4.189$ & $43.33$ & $1.056$ & $\textcolor{white}{.00}59\pm22\textcolor{white}{.}\pm8.0$ & $0.800$ & $\textcolor{white}{.00}70\pm28\textcolor{white}{.}\pm10\textcolor{white}{.}$ & $\textcolor{white}{0}16.1\pm9.8$ & $0.856$ & $\textcolor{white}{.00}18\pm11\textcolor{white}{.}\pm0.6$ \\\hline
$4.189$ & $526.7$ & $1.056$ & $\textcolor{white}{0}51.0\pm7.0\pm8.0$ & $0.798$ & $\textcolor{white}{.00}61\pm8.0\pm10\textcolor{white}{.}$ & $\textcolor{white}{0}15.1\pm3.6$ & $0.856$ & $\textcolor{white}{0}16.7\pm4.0\pm0.6$ \\\hline
$4.199$ & $526.0$ & $1.056$ & $\textcolor{white}{0}57.0\pm7.0\pm8.0$ & $0.773$ & $\textcolor{white}{.00}70\pm9.0\pm10\textcolor{white}{.}$ & $\textcolor{white}{0}17.1\pm3.8$ & $0.852$ & $\textcolor{white}{0}19.1\pm4.2\pm0.7$ \\\hline
$4.208$ & $54.95$ & $1.057$ & $\textcolor{white}{.00}58\pm21\textcolor{white}{.}\pm8.0$ & $0.752$ & $\textcolor{white}{.00}73\pm27\textcolor{white}{.}\pm10\textcolor{white}{.}$ & $\textcolor{white}{0}27.7\pm9.7$ & $0.831$ & $\textcolor{white}{.00}32\pm11\textcolor{white}{.}\pm1.9$ \\\hline
$4.209$ & $517.1$ & $1.057$ & $\textcolor{white}{0}84.0\pm7.0\pm9.0$ & $0.773$ & $\textcolor{white}{.0}103\pm9.0\pm11\textcolor{white}{.}$ & $\textcolor{white}{0}15.8\pm4.1$ & $0.852$ & $\textcolor{white}{0}17.5\pm4.5\pm0.7$ \\\hline
$4.217$ & $54.60$ & $1.057$ & $\textcolor{white}{.0}138\pm21\textcolor{white}{.}\pm10\textcolor{white}{.}$ & $0.744$ & $\textcolor{white}{.0}176\pm27\textcolor{white}{.}\pm12\textcolor{white}{.}$ & $\textcolor{white}{.00}46\pm11\textcolor{white}{.}$ & $0.752$ & $\textcolor{white}{.00}58\pm13\textcolor{white}{.}\pm8.7$ \\\hline
$4.219$ & $514.6$ & $1.056$ & $\textcolor{white}{.0}116\pm7.0\pm10\textcolor{white}{.}$ & $0.746$ & $\textcolor{white}{.0}147\pm9.0\pm12\textcolor{white}{.}$ & $\textcolor{white}{0}26.5\pm4.6$ & $0.748$ & $\textcolor{white}{0}33.5\pm5.8\pm5.2$ \\\hline
$4.226$ & $44.54$ & $1.056$ & $\textcolor{white}{.0}119\pm24\textcolor{white}{.}\pm9.0$ & $0.761$ & $\textcolor{white}{.0}148\pm30\textcolor{white}{.}\pm11\textcolor{white}{.}$ & $\textcolor{white}{.00}41\pm12\textcolor{white}{.}$ & $0.796$ & $\textcolor{white}{.00}48\pm14\textcolor{white}{.}\pm4.4$ \\\hline
$4.226$ & $1056.4$ & $1.056$ & $\textcolor{white}{.0}128\pm5.0\pm10\textcolor{white}{.}$ & $0.761$ & $\textcolor{white}{.0}160\pm6.0\pm12\textcolor{white}{.}$ & $\textcolor{white}{0}30.2\pm3.5$ & $0.796$ & $\textcolor{white}{0}35.9\pm4.2\pm3.3$ \\\hline
$4.236$ & $530.3$ & $1.056$ & $\textcolor{white}{.0}124\pm7.0\pm9.0$ & $0.800$ & $\textcolor{white}{.0}147\pm8.0\pm11\textcolor{white}{.}$ & $\textcolor{white}{0}20.0\pm4.2$ & $0.888$ & $\textcolor{white}{0}21.4\pm4.5\pm1.3$ \\\hline
$4.242$ & $55.88$ & $1.056$ & $\textcolor{white}{.0}137\pm22\textcolor{white}{.}\pm10\textcolor{white}{.}$ & $0.828$ & $\textcolor{white}{.0}156\pm25\textcolor{white}{.}\pm11\textcolor{white}{.}$ & $\textcolor{white}{0}16.3\pm8.4$ & $0.877$ & $\textcolor{white}{0}17.6\pm9.1\pm1.0$ \\\hline
$4.244$ & $538.1$ & $1.056$ & $\textcolor{white}{.0}105\pm7.0\pm9.0$ & $0.838$ & $\textcolor{white}{.0}118\pm8.0\pm10\textcolor{white}{.}$ & $\textcolor{white}{0}19.1\pm4.4$ & $0.870$ & $\textcolor{white}{0}20.8\pm4.8\pm1.0$ \\\hline
$4.258$ & $828.4$ & $1.054$ & $\textcolor{white}{0}99.0\pm6.0\pm8.0$ & $0.868$ & $108.0\pm6.0\pm9.0$ & $\textcolor{white}{00}2.1\pm7.7$ & $0.844$ & $\textcolor{white}{00}2.3\pm8.7\pm0.1$ \\\hline
$4.267$ & $531.1$ & $1.053$ & $\textcolor{white}{0}85.0\pm7.0\pm8.0$ & $0.873$ & $\textcolor{white}{0}93.0\pm7.0\pm9.0$ & $\textcolor{white}{.00}10\pm14\textcolor{white}{.}$ & $0.833$ & $\textcolor{white}{.00}11\pm16\textcolor{white}{.}\pm0.3$ \\\hline
$4.278$ & $175.7$ & $1.053$ & $\textcolor{white}{.0}107\pm12\textcolor{white}{.}\pm9.0$ & $0.882$ & $\textcolor{white}{.0}115\pm13\textcolor{white}{.}\pm9.0$ & $\textcolor{white}{.00}49\pm25\textcolor{white}{.}$ & $0.822$ & $\textcolor{white}{.00}57\pm29\textcolor{white}{.}\pm1.5$ \\\hline
$4.288$ & $502.4$ & $1.053$ & $\textcolor{white}{0}80.0\pm7.0\pm8.0$ & $0.899$ & $\textcolor{white}{0}85.0\pm8.0\pm8.0$ & $\textcolor{white}{.00}42\pm13\textcolor{white}{.}$ & $0.812$ & $\textcolor{white}{.00}49\pm15\textcolor{white}{.}\pm1.3$ \\\hline
$4.308$ & $45.08$ & $1.052$ & $\textcolor{white}{.0}125\pm23\textcolor{white}{.}\pm9.0$ & $0.950$ & $\textcolor{white}{.0}125\pm23\textcolor{white}{.}\pm9.0$ & $\textcolor{white}{.00}62\pm13\textcolor{white}{.}$ & $0.796$ & $\textcolor{white}{.00}74\pm16\textcolor{white}{.}\pm1.4$ \\\hline
$4.313$ & $501.2$ & $1.052$ & $\textcolor{white}{0}57.0\pm7.0\pm8.0$ & $0.966$ & $\textcolor{white}{0}57.0\pm7.0\pm8.0$ & $\textcolor{white}{0}60.4\pm6.7$ & $0.792$ & $\textcolor{white}{0}72.4\pm8.0\pm1.4$ \\\hline
$4.338$ & $505.0$ & $1.051$ & $\textcolor{white}{0}56.0\pm7.0\pm7.0$ & $1.049$ & $\textcolor{white}{0}51.0\pm7.0\pm7.0$ & $\textcolor{white}{0}78.1\pm6.4$ & $0.781$ & $\textcolor{white}{0}95.1\pm7.8\pm1.5$ \\\hline
$4.358$ & $544.0$ & $1.051$ & $\textcolor{white}{0}43.0\pm7.0\pm7.0$ & $0.950$ & $\textcolor{white}{0}43.0\pm7.0\pm7.0$ & $\textcolor{white}{0}92.0\pm5.4$ & $0.796$ & $110.0\pm6.5\pm2.1$ \\\hline
$4.378$ & $522.7$ & $1.051$ & $\textcolor{white}{0}28.0\pm7.0\pm7.0$ & $1.155$ & $\textcolor{white}{0}23.0\pm6.0\pm6.0$ & $100.8\pm5.9$ & $0.819$ & $117.1\pm6.9\pm2.5$ \\\hline
$4.387$ & $55.57$ & $1.051$ & $\textcolor{white}{.00}29\pm21\textcolor{white}{.}\pm7.0$ & $1.173$ & $\textcolor{white}{.00}23\pm17\textcolor{white}{.}\pm6.0$ & $\textcolor{white}{.00}83\pm13\textcolor{white}{.}$ & $0.843$ & $\textcolor{white}{.00}94\pm15\textcolor{white}{.}\pm1.4$ \\\hline
$4.397$ & $507.8$ & $1.051$ & $\textcolor{white}{0}19.0\pm7.0\pm7.0$ & $1.188$ & $\textcolor{white}{0}15.0\pm6.0\pm5.0$ & $\textcolor{white}{0}87.0\pm5.6$ & $0.871$ & $\textcolor{white}{0}95.0\pm6.1\pm1.4$ \\\hline
$4.416$ & $46.80$ & $1.052$ & $\textcolor{white}{.00}49\pm22\textcolor{white}{.}\pm6.0$ & $1.210$ & $\textcolor{white}{.00}38\pm17\textcolor{white}{.}\pm6.0$ & $\textcolor{white}{.00}52\pm13\textcolor{white}{.}$ & $0.935$ & $\textcolor{white}{.00}53\pm13\textcolor{white}{.}\pm0.7$ \\\hline
$4.416$ & $1043.0$ & $1.052$ & $\textcolor{white}{0}38.0\pm5.0\pm6.0$ & $1.210$ & $\textcolor{white}{0}29.7\pm3.8\pm5.8$ & $\textcolor{white}{0}72.0\pm3.9$ & $0.935$ & $\textcolor{white}{0}73.1\pm4.0\pm1.0$ \\\hline
$4.437$ & $569.9$ & $1.054$ & $\textcolor{white}{0}27.0\pm7.0\pm6.0$ & $1.226$ & $\textcolor{white}{0}21.0\pm5.0\pm5.0$ & $\textcolor{white}{0}61.0\pm4.5$ & $1.009$ & $\textcolor{white}{0}57.4\pm4.2\pm0.9$ \\\hline
$4.467$ & $111.1$ & $1.055$ & $\textcolor{white}{.00}32\pm15\textcolor{white}{.}\pm6.0$ & $1.235$ & $\textcolor{white}{.00}24\pm11\textcolor{white}{.}\pm5.0$ & $\textcolor{white}{0}17.8\pm6.4$ & $1.095$ & $\textcolor{white}{0}15.4\pm5.5\pm0.4$ \\\hline
$4.527$ & $112.1$ & $1.054$ & $\textcolor{white}{.00}17\pm15\textcolor{white}{.}\pm4.0$ & $1.228$ & $\textcolor{white}{.00}13\pm11\textcolor{white}{.}\pm3.2$ & $\textcolor{white}{0}21.5\pm6.0$ & $1.177$ & $\textcolor{white}{0}17.3\pm4.9\pm0.5$ \\\hline
$4.575$ & $48.93$ & $1.054$ & $\textcolor{white}{.00}17\pm23\textcolor{white}{.}\pm4.0$ & $1.214$ & $\textcolor{white}{.00}14\pm18\textcolor{white}{.}\pm3.5$ & $\textcolor{white}{0}13.5\pm9.2$ & $1.154$ & $\textcolor{white}{0}11.1\pm7.5\pm0.5$ \\\hline
$4.600$ & $586.9$ & $1.055$ & $\textcolor{white}{0}28.0\pm6.0\pm4.0$ & $1.205$ & $\textcolor{white}{0}22.0\pm5.0\pm3.4$ & $\textcolor{white}{0}21.6\pm2.9$ & $1.106$ & $\textcolor{white}{0}18.5\pm2.5\pm0.5$ \\\hline
$4.612$ & $103.8$ & $1.055$ & $\textcolor{white}{.00}30\pm15\textcolor{white}{.}\pm4.0$ & $1.201$ & $\textcolor{white}{.00}23\pm12\textcolor{white}{.}\pm3.7$ & $\textcolor{white}{0}25.8\pm6.9$ & $1.071$ & $\textcolor{white}{0}22.8\pm6.1\pm0.9$ \\\hline
$4.628$ & $521.5$ & $1.054$ & $\textcolor{white}{0}10.3\pm7.1\pm3.9$ & $1.195$ & $\textcolor{white}{00}8.2\pm5.6\pm3.1$ & $\textcolor{white}{0}30.2\pm3.4$ & $1.017$ & $\textcolor{white}{0}28.1\pm3.2\pm2.3$ \\\hline
$4.641$ & $552.4$ & $1.054$ & $\textcolor{white}{00}3.0\pm7.0\pm4.0$ & $1.191$ & $\textcolor{white}{00}2.0\pm5.6\pm3.2$ & $\textcolor{white}{0}30.7\pm3.3$ & $0.976$ & $\textcolor{white}{0}29.8\pm3.2\pm3.5$ \\\hline
$4.661$ & $529.6$ & $1.054$ & $\textcolor{white}{0}14.0\pm7.0\pm4.0$ & $1.184$ & $\textcolor{white}{0}10.8\pm5.7\pm3.2$ & $\textcolor{white}{0}31.1\pm3.4$ & $0.949$ & $\textcolor{white}{0}31.1\pm3.4\pm4.2$ \\\hline
$4.682$ & $1669.3$ & $1.054$ & $\textcolor{white}{0}17.9\pm5.0\pm3.8$ & $1.178$ & $\textcolor{white}{0}14.5\pm4.1\pm3.2$ & $\textcolor{white}{0}32.8\pm2.3$ & $0.996$ & $\textcolor{white}{0}31.2\pm2.2\pm2.6$ \\\hline
$4.699$ & $536.5$ & $1.055$ & $\textcolor{white}{0}14.4\pm7.1\pm3.7$ & $1.172$ & $\textcolor{white}{0}11.6\pm5.7\pm3.0$ & $\textcolor{white}{0}26.3\pm3.2$ & $1.064$ & $\textcolor{white}{0}23.4\pm2.9\pm0.4$ \\\hline
$4.740$ & $164.3$ & $1.055$ & $\textcolor{white}{.00}20\pm13\textcolor{white}{.}\pm3.4$ & $1.161$ & $\textcolor{white}{.00}17\pm10\textcolor{white}{.}\pm2.9$ & $\textcolor{white}{0}23.6\pm5.3$ & $1.180$ & $\textcolor{white}{0}19.0\pm4.3\pm2.2$ \\\hline
$4.750$ & $367.2$ & $1.055$ & $\textcolor{white}{0}12.3\pm8.6\pm3.3$ & $1.158$ & $\textcolor{white}{0}10.1\pm7.0\pm2.8$ & $\textcolor{white}{0}19.4\pm3.5$ & $1.193$ & $\textcolor{white}{0}15.4\pm2.8\pm2.0$ \\\hline
$4.781$ & $512.8$ & $1.055$ & $\textcolor{white}{0}37.3\pm7.2\pm3.9$ & $1.150$ & $\textcolor{white}{0}30.8\pm6.0\pm3.6$ & $\textcolor{white}{0}19.9\pm3.1$ & $1.211$ & $\textcolor{white}{0}15.6\pm2.4\pm2.5$ \\\hline
$4.843$ & $527.3$ & $1.056$ & $\textcolor{white}{0}12.3\pm7.3\pm3.1$ & $1.136$ & $\textcolor{white}{0}10.3\pm6.1\pm2.6$ & $\textcolor{white}{0}19.7\pm3.0$ & $1.204$ & $\textcolor{white}{0}15.5\pm2.4\pm2.5$ \\\hline
$4.918$ & $208.1$ & $1.056$ & $\textcolor{white}{.00}19\pm12\textcolor{white}{.}\pm3.0$ & $1.122$ & $\textcolor{white}{0}15.9\pm9.9\pm2.7$ & $\textcolor{white}{0}15.9\pm4.5$ & $1.181$ & $\textcolor{white}{0}12.8\pm3.6\pm1.9$ \\\hline
$4.951$ & $160.4$ & $1.056$ & $\textcolor{white}{.00}13\pm14\textcolor{white}{.}\pm2.7$ & $1.117$ & $\textcolor{white}{.00}11\pm12\textcolor{white}{.}\pm2.3$ & $\textcolor{white}{0}22.7\pm5.3$ & $1.170$ & $\textcolor{white}{0}18.4\pm4.3\pm2.7$ \\\hline\hline
  
\end{tabular}
\end{center}
\label{CS}
\end{table*}

Charged tracks detected in the MDC are required to satisfy $|\cos\theta| < 0.93$; the polar angle $\theta$ is defined with respect to the $z$-axis, 
which is the symmetry axis of the MDC. The distance of closest approach to the interaction point (IP) for charged tracks must be less than 10\,cm along the $z$-axis, $|V_{z}|$,  and less than 1.0\,cm in the transverse plane, $|V_{xy}|$.

Photon candidates are identified using showers in the EMC.  The deposited energy of each shower must be more than 25~MeV in the barrel region ($|\cos \theta|< 0.80$) and more than 50~MeV in the end cap region ($0.86 <|\cos \theta|< 0.92$). To exclude showers that originate from charged tracks, the angle subtended by the EMC shower and the position of the closest charged track at the EMC must be greater than 20~degrees. To suppress electronic noise and showers unrelated to the event, the difference between the EMC time and the event start time is required to be within [0, 700]\,ns. 

The topology of events allows one to separate inclusive $J/\psi$ and $\psi(3686)$ events from those with the ISR return to the $J/\psi$ and $\psi(3686)$ resonances. Table~\ref{TRK} shows track and photon configurations that must be present in the candidate event for each process under consideration. Candidate events with the $J/\psi$ meson in the final state must contain at least one pair of oppositely-charged lepton tracks. For muon and electron identification, the energy deposited by a charged track in the EMC must be less than $0.6$ and more than $0.9$~GeV, respectively. The di-lepton invariant masses should fall within the range of $(2.8, 3.4)$~GeV/$c^{2}$.

\begin{table*}[htbp]
\caption{Track and photon configurations that must be contained in the candidate event for each process under consideration. The corresponding observables from Eqns.~(\ref{CS_JpsiX_prompt})$-$(\ref{CS_psiX_prompt}) are indicated in the second column. Positive- and negative-charged tracks are denoted by~``$\oplus$" and ``$\ominus$",  respectively. The notation~``$\forall$"  means that the candidate event can contain any number of photons.}
\begin{center}
\begin{tabular}{cccc}\hline\hline
Process & Observable & Charged tracks & Photons  \\\hline
$e^{+}e^{-}\rightarrow \gamma_{\textmd{ISR}}J/\psi$ & $N_{\gamma_{\textmd{ISR}}J/\psi }$ & $1\oplus$ and $1\ominus$ & $0$ or $1$  \\\hline
$e^{+}e^{-}\rightarrow J/\psi X$ & $N_{J/\psi X}$ & $1\oplus$ and $1\ominus$ & $\geq 2$  \\\hline
$e^{+}e^{-}\rightarrow J/\psi X$ & $N_{J/\psi X}$ & $>1\oplus$ or $>1\ominus$ &  $\forall$ \\\hline
$e^{+}e^{-}\rightarrow \gamma_{\textmd{ISR}}\psi(3686)$ & $N^{l}_{\gamma_{\textmd{ISR}}\psi }$ & $2\oplus$ and $2\ominus$ & $0$ or $1$  \\\hline
$e^{+}e^{-}\rightarrow \psi(3686) X$ & $N^{l}_{\psi X}$ & $2\oplus$ and $2\ominus$ & $\geq 2$  \\\hline
$e^{+}e^{-}\rightarrow \psi(3686) X$ & $N^{l}_{\psi X}$ & $>2\oplus$ or $>2\ominus$ &  $\forall$ \\\hline
$e^{+}e^{-}\rightarrow  \chi_{cJ} X$ & $N_{\chi_{cJ} X}$ & $1\oplus$ and $1\ominus$ & $\geq 2$  \\\hline
$e^{+}e^{-}\rightarrow  \chi_{cJ} X$ & $N_{\chi_{cJ} X}$ & $>1\oplus$ or $>1\ominus$ & $\geq 1$  \\\hline\hline
\end{tabular}
\end{center}
\label{TRK}
\end{table*}

To improve the resolution of the invariant mass in the reconstruction of $\psi(3686)$ and $\chi_{cJ}$~(\textit{J} = 1, 2) mesons, a kinematic fit is performed by constraining the di-lepton invariant mass to be the nominal $J/\psi$ mass.  After this fit we choose only the pair of leptons with the minimum fit $\chi^{2}$. Events with large values of the minimum $\chi^{2}$ are excluded from further consideration. The momenta updated by the kinematic fit are used in subsequent $\psi(3686)$ and $\chi_{cJ}$ meson reconstruction. Candidate events with the $\psi(3686)$ or $\chi_{cJ}$ meson in the final state must contain at least four or two charged tracks with zero net charge, respectively. Moreover, there must be a pair of oppositely-charged tracks satisfying the $J/\psi$ selection criteria described above. For the $\psi(3686)$ meson, any other pairs of oppositely-charged tracks are considered as $\pi^{+}\pi^{-}$ without any particle identification.

\begin{figure*}
\begin{center}
\includegraphics[width=1.\textwidth]{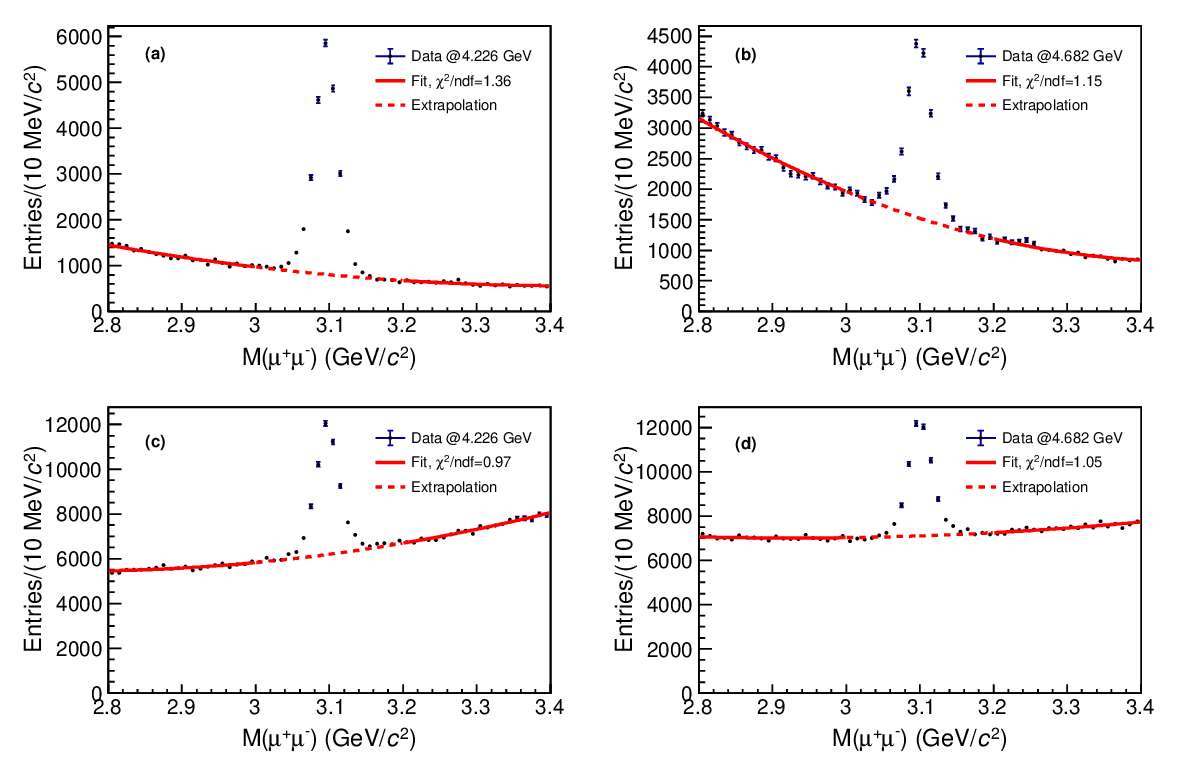}
\caption{Di-muon invariant mass spectra for $e^{+}e^{-} \rightarrow J/\psi X$~(a, b) and $e^{+}e^{-}\rightarrow \gamma_{\textmd{ISR}}J/\psi$~(c, d) events at center-of-mass energies 4.226~(a, c) and 4.682~(b, d)~GeV. The fitting function and its extrapolation into the signal region are shown in red by the solid and dotted lines, respectively.}
\label{fig:N_Jpsi}
\end{center}
\end{figure*}

For all processes under consideration, the number of observed signal events at each center-of-mass energy point is calculated as the difference between the count of total events and the integral of the fitted background functions in the signal region. The statistical uncertainty of the signal yield is taken as the quadratic sum of the statistical uncertainties of the corresponding number of total and background events. To estimate the number of background events in the signal region for $e^{+}e^{-} \rightarrow J/\psi X$ and $e^{+}e^{-}\rightarrow \gamma_{\textmd{ISR}}J/\psi$ processes, a quadratic function is fitted to a smooth background part of the di-muon invariant mass spectrum in the range $(2.8, 3.4)$~GeV/$c^{2}$, excluding the nominal $J/\psi$ mass region between $(3.0, 3.2)$~GeV/$c^2$, i.e., the signal region. For all data samples, the minimum statistical uncertainty of the number of $e^{+}e^{-} \rightarrow J/\psi X$ events is an order of magnitude greater than the maximum number of events for which two di-muon combinations fall into the signal region. Figure~\ref{fig:N_Jpsi} shows the di-muon invariant mass spectra, as well as the fitting function and its extrapolation into the signal region, for the data samples with center-of-mass energies of 4.226 and 4.682~GeV. The observed numbers of $e^{+}e^{-} \rightarrow J/\psi X$ events at $\sqrt{s}=4.226$ and $4.682$~GeV are $20~277\pm214$ and $12~697\pm315$, respectively. Similarly, the observed numbers of $e^{+}e^{-}\rightarrow \gamma_{\textmd{ISR}}J/\psi$ events at $\sqrt{s}=4.226$ and $4.682$~GeV are $24~586\pm472$ and $21~563\pm511$, respectively.

\begin{figure*}
\begin{center}
\includegraphics[width=1.\textwidth]{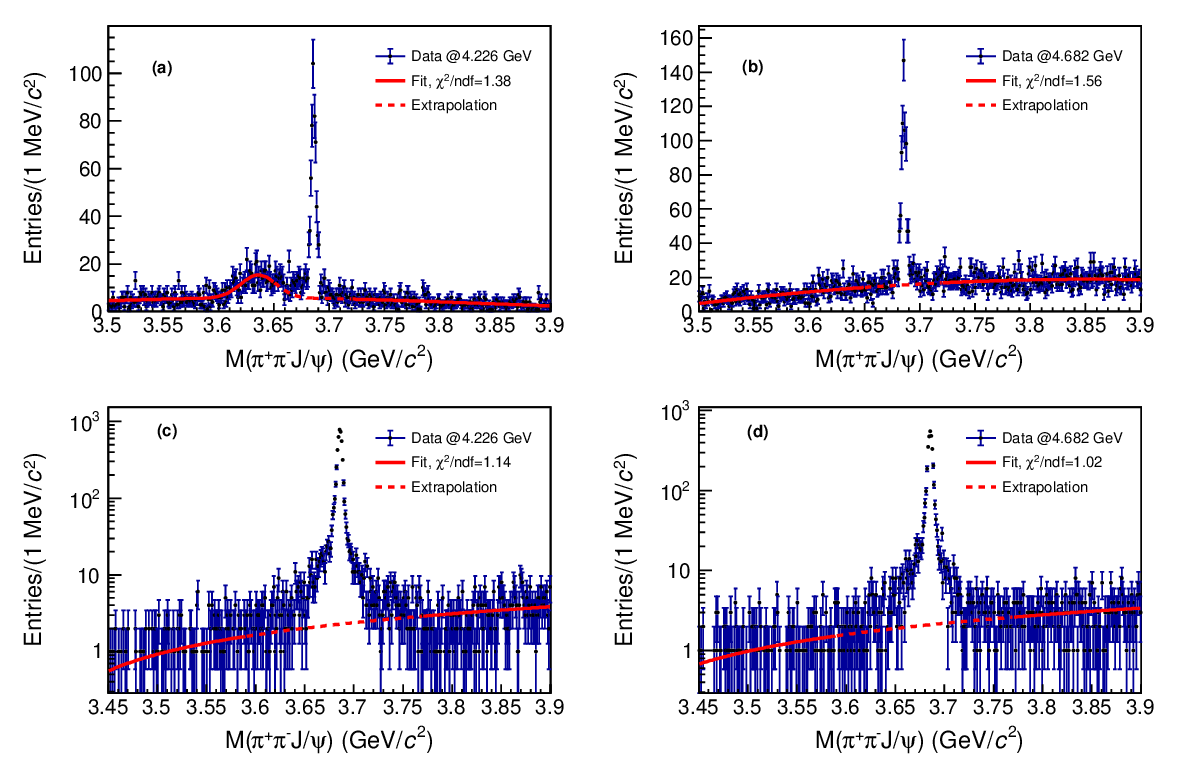}
\caption{The $\pi^+\pi^-J/\psi$ invariant mass spectra for $e^{+}e^{-}\rightarrow  \psi(3686) X$~(a, b) and $e^{+}e^{-}\rightarrow \gamma_{\textmd{ISR}}\psi(3686)$~(c, d) events at center-of-mass energies 4.226~(a, c) and 4.682~(b, d)~GeV. The fitting function and its extrapolation into the signal region are shown in red by the solid and dotted lines, respectively. The $J/\psi \rightarrow \mu^{+}\mu^{-}$ decay mode is used to reconstruct events.}
\label{fig:N_psi}
\end{center}
\end{figure*}

\begin{figure*}
\begin{center}
\includegraphics[width=1.\textwidth]{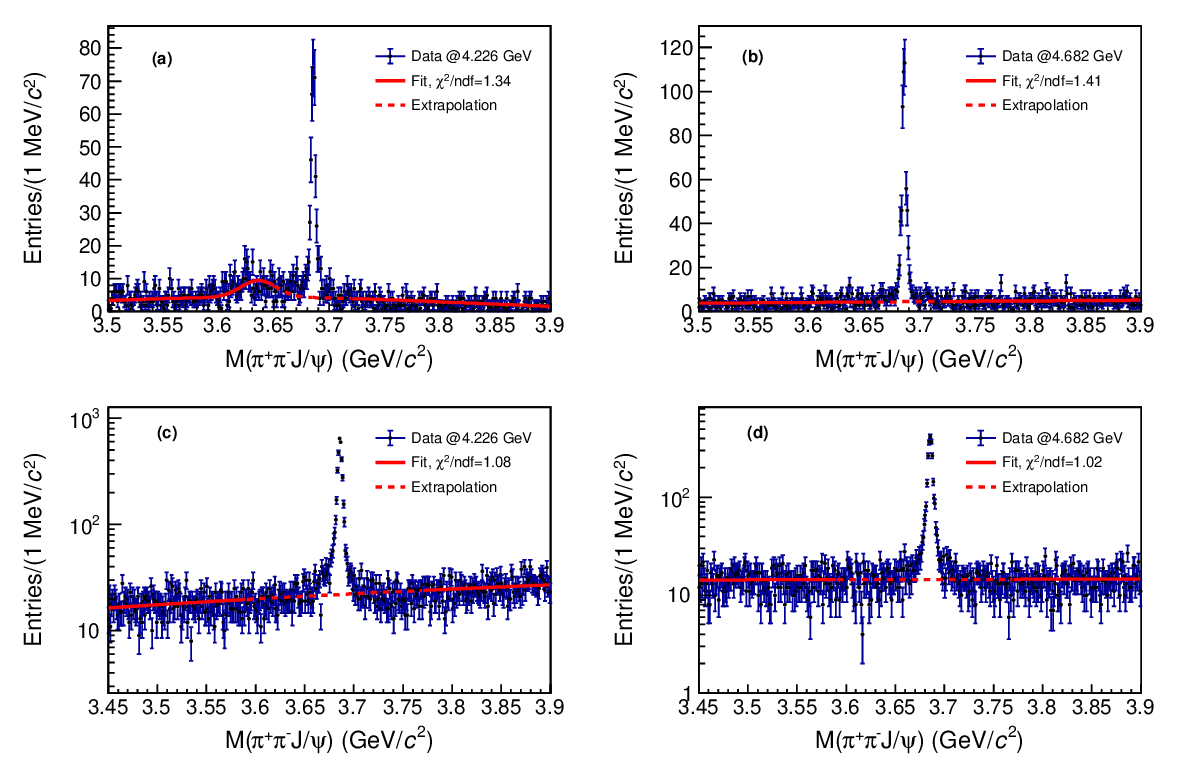}
\caption{The $\pi^+\pi^-J/\psi$ invariant mass spectra for $e^{+}e^{-}\rightarrow  \psi(3686) X$~(a, b) and $e^{+}e^{-}\rightarrow \gamma_{\textmd{ISR}}\psi(3686)$~(c, d) events at center-of-mass energies 4.226~(a, c) and 4.682~(b, d)~GeV. The fitting function and its extrapolation into the signal region are shown in red by the solid and dotted lines, respectively. The $J/\psi \rightarrow e^{+}e^{-}$ decay mode is used to reconstruct events.}
\label{fig:N_psi_ee}
\end{center}
\end{figure*}

To estimate the number of background events in the signal region for the $e^{+}e^{-}\rightarrow  \psi(3686) X$ process, a linear function is fitted to the smooth background part of the $\pi^{+}\pi^{-} J/\psi$ invariant mass distribution in the range $(3.50, 3.90)$~GeV/$c^{2}$, excluding the nominal $\psi(3686)$ signal region between $(3.65, 3.72)$~GeV/$c^{2}$. However, for energy points with $\sqrt{s}\geq 4.600$~GeV, a quadratic function is used instead of a linear one due to the distinctive background shapes observed at these points in the di-muon channel. Additionally, for energy points with high integrated luminosity in the range from $\sqrt{s}=4.129$ to $4.437$~GeV, a double Gaussian curve substitutes the linear function. This Gaussian curve characterizes a peak, which comes from the predominant exclusive process $e^{+}e^{-}\rightarrow \pi^{+}\pi^{-}\psi(3686), \psi(3686) \rightarrow \pi^{+}\pi^{-} J/\psi$, with the $\pi^{+}\pi^{-} J/\psi$ invariant mass including pions not originating from the $\psi(3686)$ decay. The location of this peak depends on the collision energy, and one of the Gaussian function widths is fixed according to the phase space signal MC samples at each energy point. The second Gaussian describes a broad background component. To estimate the number of background events in the signal region for the $e^{+}e^{-}\rightarrow  \gamma_{\textmd{ISR}}\psi(3686)$ process, a linear function is fitted to a smooth background part of the $\pi^{+}\pi^{-} J/\psi$ invariant mass distribution in the range of $(3.45, 3.90)$~GeV/$c^{2}$, excluding the nominal $\psi(3686)$ mass region between $(3.59, 3.77)$~GeV/$c^{2}$. Figures~\ref{fig:N_psi} and~\ref{fig:N_psi_ee} show the $\pi^+\pi^-J/\psi$ invariant mass spectra in the di-muon and di-electron channels, respectively, for the data samples with center-of-mass energies of 4.226 and 4.682~GeV. The peak, which comes from the exclusive process $e^{+}e^{-}\rightarrow \pi^{+}\pi^{-}\psi(3686) \rightarrow \pi^{+}\pi^{-}(\pi^{+}\pi^{-} J/\psi)$, can be found at an invariant mass of 3.636 GeV/$c^2$ in Fig.~\ref{fig:N_psi}~(a) and Fig.~\ref{fig:N_psi_ee}~(a). For the di-muon decay mode, the observed numbers of $e^{+}e^{-} \rightarrow \psi(3686) X$ events at $\sqrt{s}=4.226$ and $4.682$~GeV are $693\pm37$ and $733\pm45$, respectively. For the di-electron decay mode,  the observed numbers of $e^{+}e^{-} \rightarrow \psi(3686) X$ events at $\sqrt{s}=4.226$ and $4.682$~GeV are $454\pm31$ and $648\pm32$, respectively. Similarly, for the di-muon decay mode, the observed numbers of $e^{+}e^{-}\rightarrow \gamma_{\textmd{ISR}}\psi(3686)$ events at $\sqrt{s}=4.226$ and $4.682$~GeV are $5~174\pm76$ and $3~565\pm64$, respectively. For the di-electron decay mode, the observed numbers of $e^{+}e^{-}\rightarrow \gamma_{\textmd{ISR}}\psi(3686)$ events at $\sqrt{s}=4.226$ and $4.682$~GeV are $3~496\pm100$ and $2~486\pm83$, respectively.

\begin{figure*}
\begin{center}
\includegraphics[width=1.\textwidth]{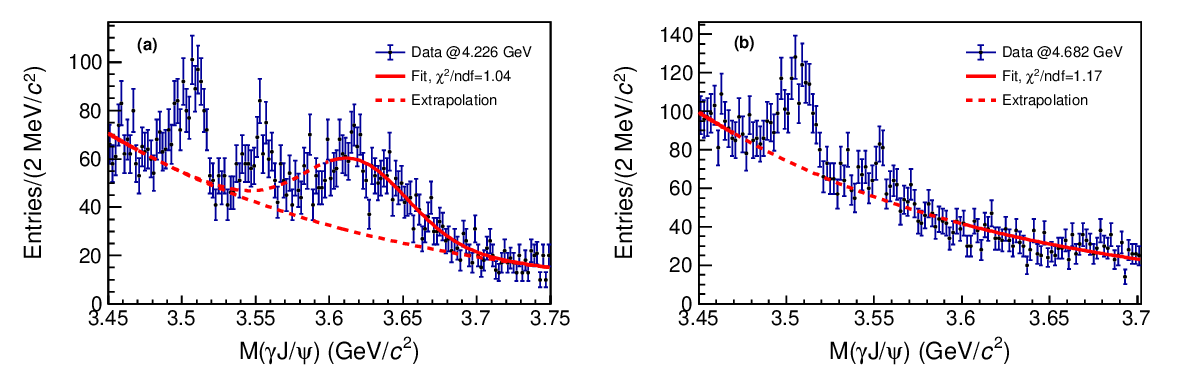}
\caption{The $\gamma J/\psi$ invariant mass spectra for $e^{+}e^{-}\rightarrow  \chi_{cJ} X$~(\textit{J} = 1, 2) events at center-of-mass energies 4.226~(a) and 4.682~(b)~GeV. The fitting function and its extrapolation into the signal region are shown in red by the solid and dotted lines, respectively.}
\label{fig:N_chic}
\end{center}
\end{figure*}

To estimate the number of background events in the signal region for the $e^{+}e^{-}\rightarrow  \chi_{cJ} X$~(\textit{J} = 1, 2) processes, the sum of an exponential and a Gaussian functions is used to describe the background in the $\gamma J/\psi$ invariant mass distribution. This fitting procedure covers the range of $(3.45, 3.70)$~GeV/$c^{2}$, excluding the signal regions $(3.47, 3.53)$~GeV/$c^2$ and $(3.53, 3.60)$~GeV/$c^2$ around the nominal $\chi_{c1}$ and $\chi_{c2}$ masses, respectively. However, for energy points in the range from $\sqrt{s}=4.085$ to $4.397$~GeV, the Gaussian curve accounts for a peak originating from the process $e^{+}e^{-}\rightarrow \gamma_{\textmd{ISR}}\psi(3686)$, where the $\gamma J/\psi$ invariant mass includes the ISR photon. The position of such a peak depends on the collision energy. The means and widths of the Gaussian functions are determined according to the {\sc kkmc} samples $e^{+}e^{-}\rightarrow \gamma_{\textmd{ISR}}\psi(3686)$ with $\psi(3686)  \rightarrow \pi^{+}\pi^{-}J/\psi$ and $J/\psi \rightarrow \mu^{+}\mu^{-}$  decay modes at each energy point. Figure~\ref{fig:N_chic} shows the $\gamma J/\psi$ invariant mass spectra for the data samples with center-of-mass energies of 4.226 and 4.682~GeV. The peak, which comes from the process $e^{+}e^{-}\rightarrow \gamma_{\textmd{ISR}}\psi(3686)$, is visible in Fig.~\ref{fig:N_chic}~(a) for a $\gamma J/\psi$ invariant mass of 3.618~GeV/$c^2$. Moreover, for the data sample with $\sqrt{s}=4.226$~GeV, the range of the $\gamma J/\psi$ invariant mass distribution was slightly expanded, to $(3.45, 3.75)$~GeV/$c^{2}$, in order to improve the description of the background peak. The observed numbers of $e^{+}e^{-} \rightarrow \chi_{c1} X$ events at $\sqrt{s}=4.226$ and $4.682$~GeV are $424\pm76$ and $557\pm84$, respectively. Similarly, the observed numbers of $e^{+}e^{-}\rightarrow \chi_{c2} X$ events at $\sqrt{s}=4.226$ and $4.682$~GeV are $156\pm54$ and $152\pm55$, respectively.

\section{EFFICIENCY}

The reconstruction efficiencies reflect the geometrical acceptance of the detector, particle detection and reconstruction efficiencies, and the event selection procedures. To calculate the reconstruction efficiency for the particular process,  the number of reconstructed MC events in the signal region of the corresponding invariant mass distribution is determined.  

Since the total $J/\psi$ yield predominantly consists of the ISR return to the $\psi(3686)$ resonance, in order to determine reconstruction efficiency for the $J/\psi$ meson in the process $e^{+}e^{-}\rightarrow J/\psi X$, we produce {\sc kkmc} samples $e^{+}e^{-}\rightarrow \gamma_{\textmd{ISR}}\psi(3686)$ with the following major $\psi(3686)$ to $J/\psi$ decay modes: $\psi(3686)  \rightarrow \pi^{+}\pi^{-}J/\psi$, $\psi(3686)  \rightarrow \pi^{0}\pi^{0}J/\psi$, $\psi(3686)  \rightarrow \eta J/\psi$, $\psi(3686)  \rightarrow \gamma \chi_{c1},\chi_{c1} \rightarrow \gamma J/\psi$, $\psi(3686)  \rightarrow \gamma \chi_{c2}, \chi_{c2} \rightarrow \gamma J/\psi$, all with $J/\psi \rightarrow \mu^{+}\mu^{-}$. The combined branching fraction of these selected channels agrees with the branching fraction of the $\psi(3686)  \rightarrow J/\psi  X$  decay within the uncertainty. To ensure the adequacy of the MC simulation of the $e^{+}e^{-}\rightarrow J/\psi X$ process, we perform an additional study using the inclusive MC samples. We compare the reconstruction efficiency for the $J/\psi$ meson in the process $e^{+}e^{-}\rightarrow J/\psi X$ using the following subsamples of the inclusive MC samples: $e^{+}e^{-}\rightarrow \gamma_{\textmd{ISR}}\psi(3686)$,  $e^{+}e^{-}\rightarrow \psi(3686) X$, all events with $\psi(3686)$ mesons, and $e^{+}e^{-}\rightarrow J/\psi X$ decays. For the first three subsamples, we select events with $\psi(3686) \rightarrow J/\psi X$ and for all four we select $J/\psi \rightarrow \mu^{+}\mu^{-}$  decay modes. We find that the addition of $e^{+}e^{-}\rightarrow \psi(3686) X$  channels, as well as hidden $e^{+}e^{-}\rightarrow J/\psi X$ channels, leads to a change in efficiency within a systematic uncertainty of 2\%. The comparison of the momentum and angular distributions for the $J/\psi$, as well as charged and neutral tracks, in data and MC is also satisfactory. For all MC samples, the number of events for which two di-muon combinations fall into the signal region is negligible. The reconstruction efficiencies for the $J/\psi$ meson in the process $e^{+}e^{-}\rightarrow J/\psi X$ at $\sqrt{s}=4.226$ and $4.682$~GeV are $(73.8\pm1.5)\%$ and $(71.7\pm1.4)\%$, respectively.  Systematic uncertainties of the reconstruction efficiencies are indicated here and further.  More detail about the assessment of systematic uncertainties is given in the corresponding section.

To estimate the percentage $\mathcal{R}_{\gamma_{\textmd{ISR}}J/\psi}$ of the background $e^{+}e^{-}\rightarrow \gamma_{\textmd{ISR}}J/\psi$  events in the overall number of observed $e^{+}e^{-}\rightarrow J/\psi X$ events, we produce {\sc kkmc} samples $e^{+}e^{-}\rightarrow \gamma_{\textmd{ISR}}J/\psi$ with $J/\psi \rightarrow \mu^{+}\mu^{-}$  decay mode. The following formula is used to calculate the coefficient $\mathcal{R}_{\gamma_{\textmd{ISR}}J/\psi} $:
\begin{equation}
  \mathcal{R}_{\gamma_{\textmd{ISR}}J/\psi} = \frac{N^{\textmd{MC}}_{J/\psi X}}{N^{\textmd{MC}}_{\gamma_{\textmd{ISR}}J/\psi }},
  \label{R_JpsiISR}
\end{equation}
where $N^{\textmd{MC}}_{J/\psi X}$ denotes the number of events that satisfy the selection criteria for the process $e^{+}e^{-}\rightarrow J/\psi X$, while $N^{\textmd{MC}}_{\gamma_{\textmd{ISR}}J/\psi }$ represents the number of events that satisfy the selection criteria for the process $e^{+}e^{-}\rightarrow \gamma_{\textmd{ISR}}J/\psi$.  The coefficients $\mathcal{R}_{\gamma_{\textmd{ISR}}J/\psi}$ at $\sqrt{s}=4.226$ and $4.682$~GeV are $(2.07\pm0.08)\%$ and $(2.08\pm0.08)\%$, respectively.

To evaluate the reconstruction efficiency for the $\psi(3686)$ meson in the process $e^{+}e^{-}\rightarrow \psi(3686) X$, we produce a mixture of the major exclusive channels with typical topologies weighted according to the relative fraction of events with corresponding charged track multiplicity in the real data. For this purpose, we use a mixture of the phase space MC samples $e^{+}e^{-}\rightarrow \pi^{0}\pi^{0}\psi(3686)$ and $e^{+}e^{-}\rightarrow \pi^{+}\pi^{-}\psi(3686)$ with $\psi(3686)  \rightarrow \pi^{+}\pi^{-}J/\psi$ and $J/\psi \rightarrow l^{+}l^{-}$~($l = e, \mu$)  decay modes that correspond to multiplicities of 4 and 6 charged tracks per event, respectively. The reconstruction efficiency for the $\psi(3686)$ meson in the process $e^{+}e^{-}\rightarrow \psi(3686) X$ weakly depends on the kinematics of a specific exclusive channel, while the difference between the reconstruction efficiency for individual exclusive channels is negligible compared to the overall systematic uncertainty in efficiency quoted below. In order to exclude a possible combinatorial background, only correctly reconstructed signal events are taken into account when evaluating efficiency. For the di-muon channel, the reconstruction efficiencies for the $\psi(3686)$ meson in the process $e^{+}e^{-}\rightarrow \psi(3686) X$ at $\sqrt{s}=4.226$ and $4.682$~GeV are $(53.0\pm2.1)\%$ and $(50.5\pm2.0)\%$, respectively. For the di-electron channel, the reconstruction efficiencies for the $\psi(3686)$ meson in the process $e^{+}e^{-}\rightarrow \psi(3686) X$ at $\sqrt{s}=4.226$ and $4.682$~GeV are $(39.7\pm1.6)\%$ and $(38.0\pm1.5)\%$, respectively.

To determine the reconstruction efficiency for the $\psi(3686)$ meson in the ISR return to the $\psi(3686)$ resonance and to estimate the percentage $\mathcal{R}_{\gamma_{\textmd{ISR}}\psi}$ of the background $e^{+}e^{-}\rightarrow \gamma_{\textmd{ISR}}\psi(3686)$  events in the overall number of observed $e^{+}e^{-}\rightarrow  \psi(3686) X$ events, we produce {\sc kkmc} samples $e^{+}e^{-}\rightarrow \gamma_{\textmd{ISR}}\psi(3686)$ with $\psi(3686)  \rightarrow \pi^{+}\pi^{-}J/\psi$ and $J/\psi \rightarrow l^{+}l^{-}$~($l = e, \mu$)  decay modes. The following formula is used to calculate the coefficient $\mathcal{R}^{l}_{\gamma_{\textmd{ISR}}\psi} $:
\begin{equation}
  \mathcal{R}^{l}_{\gamma_{\textmd{ISR}}\psi} = \frac{N^{\textmd{MC}}_{\psi X, l}}{N^{\textmd{MC}}_{\gamma_{\textmd{ISR}}\psi, l}},
  \label{R_psiISR}
\end{equation}
where $N^{\textmd{MC}}_{\psi X, l}$ signifies the number of events that satisfy the selection criteria for the process $e^{+}e^{-}\rightarrow \psi(3686) X$, while $N^{\textmd{MC}}_{\gamma_{\textmd{ISR}}\psi, l}$ corresponds to the number of events that satisfy the selection criteria for the process $e^{+}e^{-}\rightarrow \gamma_{\textmd{ISR}}\psi(3686)$. For example, the reconstruction efficiencies for the $\psi(3686)$ meson in the ISR return to the $\psi(3686)$ resonance at $\sqrt{s}=4.226$ and $4.682$~GeV are $(51.2\pm2.0)\%$ and $(47.4\pm1.9)\%$, respectively.  The ratios $\mathcal{R}^{\mu}_{\gamma_{\textmd{ISR}}\psi}$  at $\sqrt{s}=4.226$ and $4.682$~GeV are $(6.1\pm0.5)\%$ and $(6.2\pm0.5)\%$, respectively. The ratios $\mathcal{R}^{e}_{\gamma_{\textmd{ISR}}\psi}$  at $\sqrt{s}=4.226$ and $4.682$~GeV are $(5.8\pm0.5)\%$ and $(6.2\pm0.5)\%$, respectively.

Since we assume that the process $e^{+}e^{-}\rightarrow \gamma_{\textmd{ISR}}\psi(3686)$ is the predominant source of $\chi_{cJ}$~(\textit{J} = 1, 2) mesons, to evaluate the reconstruction efficiency for the $\chi_{cJ}$ meson in the process $e^{+}e^{-}\rightarrow  \chi_{cJ} X$, we generate {\sc kkmc} samples $e^{+}e^{-}\rightarrow \gamma_{\textmd{ISR}}\psi(3686)$ with subsequent decays involving $\psi(3686)  \rightarrow \gamma \chi_{cJ} \rightarrow \gamma (\gamma J/\psi)$ and $J/\psi \rightarrow \mu^{+}\mu^{-}$  decay modes. In order to exclude a possible combinatorial background, only correctly reconstructed signal events are taken into account when evaluating efficiency.  The reconstruction efficiencies for the $\chi_{c1}$ meson in the process $e^{+}e^{-}\rightarrow  \chi_{c1} X$ at $\sqrt{s}=4.226$ and $4.682$~GeV are $(52.2\pm1.2)\%$ and $(50.3\pm1.1)\%$, respectively. The reconstruction efficiencies for the $\chi_{c2}$ meson in the process $e^{+}e^{-}\rightarrow  \chi_{c2} X$ at $\sqrt{s}=4.226$ and $4.682$~GeV are determined as $(43.9\pm1.0)\%$ and $(42.3\pm0.9)\%$, respectively.

\section{BACKGROUND ANALYSIS}

In order to study potential background events in the $J/\psi$ meson reconstruction, we generated {\sc kkmc} samples for $e^{+}e^{-}\rightarrow \gamma_{\textmd{ISR}}J/\psi$ with the following decay modes: $J/\psi \rightarrow \mu^{+}\mu^{-}$, $J/\psi \rightarrow  \pi^{+}\pi^{-}$, $J/\psi \rightarrow   K^{+}K^{-}$, $J/\psi \rightarrow  \pi^{+}\pi^{-}\pi^{0}$.  Backgrounds from $J/\psi \rightarrow   K^{+}K^{-}$  and $J/\psi \rightarrow  \pi^{+}\pi^{-}\pi^{0}$ are negligible.  Moreover, the $J/\psi \rightarrow   K^{+}K^{-}$ decay mode is suppressed by more than 200 times in comparison to the di-muon mode, taking into account the branching fraction and the reconstruction efficiency. The yield of $J/\psi$ mesons for the di-pion mode is suppressed more than 400 times in comparison to the di-muon mode. Thus, the di-pion impurity is less than the statistical uncertainties of the number of $J/\psi$ mesons at each energy point and is also negligible.

To estimate a possible impurity from the process $e^{+}e^{-}\rightarrow \psi(3686) X$  in a subset of $e^{+}e^{-}\rightarrow \gamma_{\textmd{ISR}}\psi(3686)$ events, we use phase space MC samples of the predominant exclusive channels $e^{+}e^{-}\rightarrow \pi^{0}\pi^{0}\psi(3686)$ and $e^{+}e^{-}\rightarrow \pi^{+}\pi^{-}\psi(3686)$. We find that such an impurity is negligible since the contributions of these exclusive channels do not exceed statistical uncertainties of the number of events for the ISR return to the $\psi(3686)$ resonance in any data sample. Besides, the values of the exclusive cross section of the process $e^{+}e^{-}\rightarrow \eta\psi(3686)$ in the relevant energy region are also negligibly small according to measurements reported in Ref.~\cite{BESIII_etaPsi}. 

In order to check the background shapes in the invariant mass distributions for all considered channels, we use the inclusive MC samples, as well as additional {\sc kkmc} samples for the processes $e^{+}e^{-}\rightarrow \gamma_{\textmd{ISR}}J/\psi$ and $e^{+}e^{-}\rightarrow \gamma_{\textmd{ISR}}\psi(3686)$. Full samples of inclusive MC simulation describe the shape of the background in data well for each channel. Also, inclusive MC subsamples with $e^{+}e^{-}\rightarrow J/\psi X$ events have only the $J/\psi$ peak in the di-muon invariant mass distribution. For inclusive MC subsamples containing only $e^{+}e^{-}\rightarrow \psi(3686) X$ events, the $\pi^{+}\pi^{-} J/\psi$ invariant mass distribution also includes a peak, which comes from the dominant exclusive process $e^{+}e^{-}\rightarrow \pi^{+}\pi^{-}\psi(3686) \rightarrow \pi^{+}\pi^{-}(\pi^{+}\pi^{-} J/\psi)$. Inclusive MC subsamples with $e^{+}e^{-}\rightarrow  \chi_{cJ} X$~(\textit{J} = 1, 2) events have only the $\chi_{c1}$ and $\chi_{c2}$ peaks in the $\gamma J/\psi$ invariant mass distribution. On the other hand, the $\gamma J/\psi$ invariant mass distribution in full samples of inclusive MC contains a peak, which comes from the process $e^{+}e^{-}\rightarrow \gamma_{\textmd{ISR}}\psi(3686)$ when the $\gamma J/\psi$ system includes the ISR photon. For the process with ISR return to the $J/\psi$ ($\psi(3686)$) resonance, inclusive MC subsamples containing only $e^{+}e^{-}\rightarrow J/\psi X$ ($e^{+}e^{-}\rightarrow \psi(3686) X$) events, as well as {\sc kkmc} samples, have only the $J/\psi$ ($\psi(3686)$) peak in the corresponding invariant mass distribution.

\section{SYSTEMATIC UNCERTAINTY}

The systematic uncertainties in the measurement of the cross sections arise from sources detailed below. The main approach to estimate the contribution of a given parameter to the systematic uncertainty of the measured cross section is to vary the corresponding parameter. The difference between the nominal and varied values of the observed (or Born) cross section is taken as corresponding contribution to systematic uncertainty. Unless otherwise specified, contributions to the systematic uncertainties for both processes under consideration, i.e., $e^{+}e^{-}\rightarrow J/\psi_{\textmd{prompt}}X$ and $e^{+}e^{-}\rightarrow \psi(3686)_{\textmd{prompt}}X$, are determined with the same approach.

The first source contributing to the systematic uncertainty is an inaccuracy in the reconstruction efficiency of charged tracks or photons.  The discrepancy between MC and the data is 1\% for charged track reconstruction efficiency and 1\% for photon reconstruction efficiency. In general,  if the reconstruction of a decay requires $n$ tracks and $m$ photons, the reconstruction efficiency varies independently by $n\%$ and $m\%$ to obtain contributions to the systematic uncertainty of the reconstruction efficiency of tracks and photons, respectively. The contributions from the track and photon reconstruction efficiencies are added in quadrature. The reconstruction efficiency contributions, which are denoted by  $\delta_{\epsilon}$, to the systematic uncertainty for all the processes considered are calculated taking into account the number of charged tracks and photons required for the reconstruction of the corresponding particle.  The effects from variations from different samples are added linearly to account for their correlation.  Since the reconstruction of the $J/\psi$ meson is necessary for each channel under consideration when measuring the cross section of the process $e^{+}e^{-}\rightarrow J/\psi_{\textmd{prompt}}X$, we vary simultaneously each reconstruction efficiency ($\epsilon_{J/\psi X}$, $\epsilon^{\mu}_{\psi X}$, $\epsilon_{\gamma_{\textmd{ISR}} \psi}$, $\epsilon_{\chi_{c1} X}$, $\epsilon_{\chi_{c2} X}$) by 2\% (for the detection of the two muons) to evaluate the $\delta_{\epsilon_{J/\psi X}}$ contribution to systematic uncertainty. To obtain $\delta_{\epsilon_{\gamma_{\textmd{ISR}} \psi}}$ and $\delta_{\epsilon^{\mu}_{\psi X}}$ contributions to the systematic uncertainty, we vary separately $\epsilon_{\gamma_{\textmd{ISR}} \psi}$ and  $\epsilon^{\mu}_{\psi X}$  by 2\% (for the detection of the remaining two pions). To calculate the total uncertainty of charged track reconstruction efficiency, we sum $\delta_{\epsilon_{J/\psi X}}$, $\delta_{\epsilon_{\gamma_{\textmd{ISR}} \psi}}$, and $\delta_{\epsilon^{\mu}_{\psi X}}$ contributions linearly. To estimate $\delta_{\epsilon_{\chi_{c1} X}}$ and $\delta_{\epsilon_{\chi_{c2} X}}$ contributions to the systematic uncertainty, we vary separately $\epsilon_{\chi_{c1} X}$ and $\epsilon_{\chi_{c2} X}$  by 1\% (for the detection of the remaining photon). To calculate the total uncertainty of photon reconstruction efficiency, we sum $\delta_{\epsilon_{\chi_{c1} X}}$ and $\delta_{\epsilon_{\chi_{c2} X}}$ contributions linearly. Then the total uncertainties of charged track and photon reconstruction efficiencies are added in quadrature to obtain the contribution of the reconstruction efficiency inaccuracies to the systematic uncertainty. For the process $e^{+}e^{-}\rightarrow \psi(3686)_{\textmd{prompt}}X$, the total systematic uncertainty of track reconstruction efficiency  is $4\%$ because the reconstruction of $\psi(3686)$ meson requires the detection of two leptons and two pions. Whereas the numerator and the denominator of the coefficient $\mathcal{R}_{\gamma_{\textmd{ISR}}J/\psi}$ ($\mathcal{R}^{l}_{\gamma_{\textmd{ISR}}\psi}$) are estimated using disjoint subsets of events with different topologies, the uncertainties of their values are independent. Since the reconstruction of two (four) charged tracks is required to evaluate both the numerator and the denominator of the coefficient $\mathcal{R}_{\gamma_{\textmd{ISR}}J/\psi}$ ($\mathcal{R}^{l}_{\gamma_{\textmd{ISR}}\psi}$),  to calculate the corresponding contribution to the systematic uncertainty, we vary the coefficient $\mathcal{R}_{\gamma_{\textmd{ISR}}J/\psi}$ ($\mathcal{R}^{l}_{\gamma_{\textmd{ISR}}\psi}$) by 4\% (8\%).

To reduce the influence of statistical fluctuations in the evaluation of uncertainties related to lepton identification and the yield determinations, we combine the data into three groups based on the energy interval ($\sqrt{s}=3.808-4.178$~GeV, $\sqrt{s}=4.189-4.338$~GeV, $\sqrt{s}=4.358-4.951$~GeV). The relative uncertainties are calculated for each group, and these relative uncertainties are used to obtain absolute uncertainties at each energy point. The contribution of  inaccuracy in the muon and electron identification is obtained by varying the maximum energy of the charged track deposited in the EMC up to $0.5$ and $0.95$~GeV, respectively. To evaluate the uncertainty in the determination of the number of $e^{+}e^{-}\rightarrow J/\psi X$ or $e^{+}e^{-}\rightarrow \gamma_{\textmd{ISR}}J/\psi$ events, we use a cubic polynomial instead of a quadratic one to describe background in the di-muon invariant mass spectrum. To estimate the contribution of uncertainty in the determination of the number of $e^{+}e^{-}\rightarrow \gamma_{\textmd{ISR}}\psi(3686)$ events, we use a quadratic polynomial instead of a linear function to characterize the background in the $\pi^{+}\pi^{-}J/\psi$ invariant mass spectrum. Similarly, to evaluate  the uncertainty in the determination of the number of $e^{+}e^{-}\rightarrow \psi(3686) X$ events,  a Gaussian curve and a quadratic polynomial are used instead of a double Gaussian function to describe background in the $\pi^{+}\pi^{-}J/\psi$ invariant mass spectrum. The contribution of inaccuracy in the determination of the number of $e^{+}e^{-}\rightarrow \chi_{cJ} $~(\textit{J} = 1, 2) events is estimated by replacing the exponential curve with the one parameterized by $p_0x^{p_1}$ to describe the background in the $\gamma J/\psi$ invariant mass spectrum.

The next source of systematic uncertainty is an inaccuracy in the ISR correction used to obtain the dressed cross section. To evaluate this contribution, an alternative fitting function is used for the interpolation between the experimental points. The discrepancy between the nominal and alternate dressed cross sections is taken as the corresponding systematic uncertainty. 

We take into account the total luminosity measurement uncertainties~\cite{BeamEnergyLuminosity2020_2021,Luminosity2011_2017}, which range from 0.4\% to 1.0\%, depending on the data set, as well as uncertainties in the branching fractions~\cite{pdg} of all charmonia decays considered in the analysis, to obtain the corresponding contributions to the systematic uncertainty at each energy point. The uncertainty of the vacuum polarization factor is taken as 0.5\% from the QED calculation~\cite{ISR_formula}. 

The contributions to systematic uncertainties for the process $e^{+}e^{-}\rightarrow J/\psi_{\textmd{prompt}} X$~at $\sqrt{s}=4.226$ and $4.682$~GeV are listed in Table~\ref{SYST_JpsiX}. Assuming all sources of uncertainties to be independent (except for the case of reconstruction efficiencies discussed in detail above), we obtain the total systematic uncertainty by summing all the contributions quadratically. Table~\ref{SYST_psiX} lists the independent and common contributions to systematic uncertainties for the process $e^{+}e^{-}\rightarrow \psi(3686)_{\textmd{prompt}} X$~at center-of-mass energies $4.226$ and $4.682$~GeV.

\begin{table*}[htbp]
\caption{Contributions to the systematic uncertainties for the process $e^{+}e^{-}\rightarrow J/\psi_{\textmd{prompt}} X$~at center-of-mass energies $4.226$ and $4.682$~GeV. The contributions denoted by ``-'' are negligibly small.}
\begin{center}
\begin{tabular}{ccc}\hline\hline
Source & $4.226$~GeV & $4.682$~GeV  \\\hline
$\epsilon_{J/\psi X}$ & $2.0\%$ & $\textcolor{white}{0}2.0\%$  \\
$\epsilon^{\mu}_{\psi X}$ & $0.3\%$ & $\textcolor{white}{0}1.8\%$ \\
$\epsilon_{\gamma_{\textmd{ISR}} \psi}$ & $4.0\%$ & $13.7\%$ \\
$\epsilon_{\chi_{c1} X}$ & $0.1\%$ & $\textcolor{white}{0}0.6\%$ \\
$\epsilon_{\chi_{c2} X}$ & $-$ & $\textcolor{white}{0}0.2\%$ \\\hline
$\mathcal{R}_{\gamma_{\textmd{ISR}}J/\psi}$ & $0.3\%$ & $\textcolor{white}{0}1.4\%$ \\
$\mathcal{R}^{\mu}_{\gamma_{\textmd{ISR}} \psi}$ & $1.0\%$ & $\textcolor{white}{0}3.2\%$ \\\hline
Identification of $\mu^{\pm}$ & $1.3\%$ & $\textcolor{white}{0}0.3\%$ \\\hline
$N_{J/\psi X}$ & $0.2\%$ & $\textcolor{white}{0}0.4\%$ \\ 
$N_{\gamma_{\textmd{ISR}} J/\psi}$ & $-$ & $-$  \\ 
$N^{\mu}_{\psi X}$ & $1.7\%$ & $\textcolor{white}{0}0.5\%$ \\ 
$N^{\mu}_{\gamma_{\textmd{ISR}} \psi}$ & $0.9\%$ & $\textcolor{white}{0}0.1\%$ \\ 
$N_{\chi_{cJ} X}$ & $1.6\%$ & $\textcolor{white}{0}4.0\%$ \\\hline
$\mathcal{L}$ & $0.7\%$ & $\textcolor{white}{0}0.5\%$ \\\hline
$\mathcal{B}_{J/\psi \rightarrow \mu^{+}\mu^{-}}$ & $0.6\%$ & $\textcolor{white}{0}0.6\%$ \\
$\mathcal{B}_{\psi  \rightarrow \pi^{+}\pi^{-} J/\psi}$ & $1.9\%$ & $\textcolor{white}{0}7.0\%$ \\
$\mathcal{\tilde B}_{\psi  \rightarrow J/\psi X}$ & $2.4\%$ & $\textcolor{white}{0}9.0\%$ \\\hline
$(1 + \delta)$ & $0.5\%$ & $\textcolor{white}{0}5.0\%$ \\
$|1-\Pi|^{-2}$ & $0.5\%$ & $\textcolor{white}{0}0.5\%$ \\\hline
Total & $8.0\%$ & $22.0\%$ \\\hline\hline
\end{tabular}
\end{center}
\label{SYST_JpsiX}
\end{table*}

\begin{table*}[htbp]
\caption{Independent and common contributions to the systematic uncertainties for the process $e^{+}e^{-}\rightarrow \psi(3686)_{\textmd{prompt}} X$~at center-of-mass energies $4.226$ and $4.682$~GeV. The contributions denoted by ``-'' are negligibly small.}
\begin{center}
\begin{tabular}{ccccc}\hline\hline
\multicolumn{1}{c}{} & 
\multicolumn{2}{c}{$J/\psi \rightarrow \mu^{+}\mu^{-}$ \; } & 
\multicolumn{2}{c}{$ J/\psi \rightarrow e^{+}e^{-}$ \; } \\
Source & $4.226$~GeV & $4.682$~GeV & $4.226$~GeV & $4.682$~GeV \\\hline
& & Independent \\\hline
Identification of $l^{\pm}$ & $\textcolor{white}{0}0.6\%$ & $\textcolor{white}{0}0.9\%$ & $\textcolor{white}{0}0.1\%$ & $\textcolor{white}{0}0.4\%$ \\
$N^{l}_{\psi X}$ & $15.0\%$ & $\textcolor{white}{0}0.8\%$ & $\textcolor{white}{0}2.0\%$ & $\textcolor{white}{0}5.0\%$ \\ 
$N^{l}_{\gamma_{\textmd{ISR}} \psi}$ & $\textcolor{white}{0}0.5\%$ & $-$  & $\textcolor{white}{0}2.0\%$ & $\textcolor{white}{0}2.0\%$ \\ 
$\mathcal{R}^{l}_{\gamma_{\textmd{ISR}} \psi}$ & $\textcolor{white}{0}7.0\%$ & $\textcolor{white}{0}3.5\%$  & $\textcolor{white}{0}6.0\%$ & $\textcolor{white}{0}2.5\%$ \\
$\epsilon^{l}_{\psi X}$ & $\textcolor{white}{0}4.0\%$ & $\textcolor{white}{0}4.0\%$ & $\textcolor{white}{0}4.0\%$ & $\textcolor{white}{0}4.0\%$ \\
$\mathcal{B}_{J/\psi \rightarrow l^{+}l^{-}}$ & $\textcolor{white}{0}0.6\%$ & $\textcolor{white}{0}0.6\%$ & $\textcolor{white}{0}0.5\%$ & $\textcolor{white}{0}0.5\%$ \\\hline
Total independent & $17.0\%$ & $\textcolor{white}{0}5.5\%$ & $\textcolor{white}{0}8.0\%$ & $\textcolor{white}{0}7.0\%$ \\\hline
& & Common \\\hline
$\mathcal{B}_{\psi  \rightarrow \pi^{+}\pi^{-} J/\psi}$ & $\textcolor{white}{0}0.9\%$ & $\textcolor{white}{0}0.9\%$  & $\textcolor{white}{0}0.9\%$ & $\textcolor{white}{0}0.9\%$ \\
$\mathcal{L}$ & $\textcolor{white}{0}0.7\%$ & $\textcolor{white}{0}0.5\%$  & $\textcolor{white}{0}0.7\%$ & $\textcolor{white}{0}0.5\%$ \\
$(1 + \delta)$ & $\textcolor{white}{0}9.0\%$ & $\textcolor{white}{0}8.0\%$ & $\textcolor{white}{0}9.0\%$ & $\textcolor{white}{0}8.0\%$ \\
$|1-\Pi|^{-2}$ & $\textcolor{white}{0}0.5\%$ & $\textcolor{white}{0}0.5\%$ & $\textcolor{white}{0}0.5\%$ & $\textcolor{white}{0}0.5\%$ \\\hline
Total common & $\textcolor{white}{0}9.0\%$ & $\textcolor{white}{0}8.0\%$ & $\textcolor{white}{0}9.0\%$ & $\textcolor{white}{0}8.0\%$ \\\hline\hline
\end{tabular}
\end{center}
\label{SYST_psiX}
\end{table*}

\section{CROSS SECTION MEASUREMENT}

\begin{figure}
\begin{center}
\includegraphics[width=0.5\textwidth]{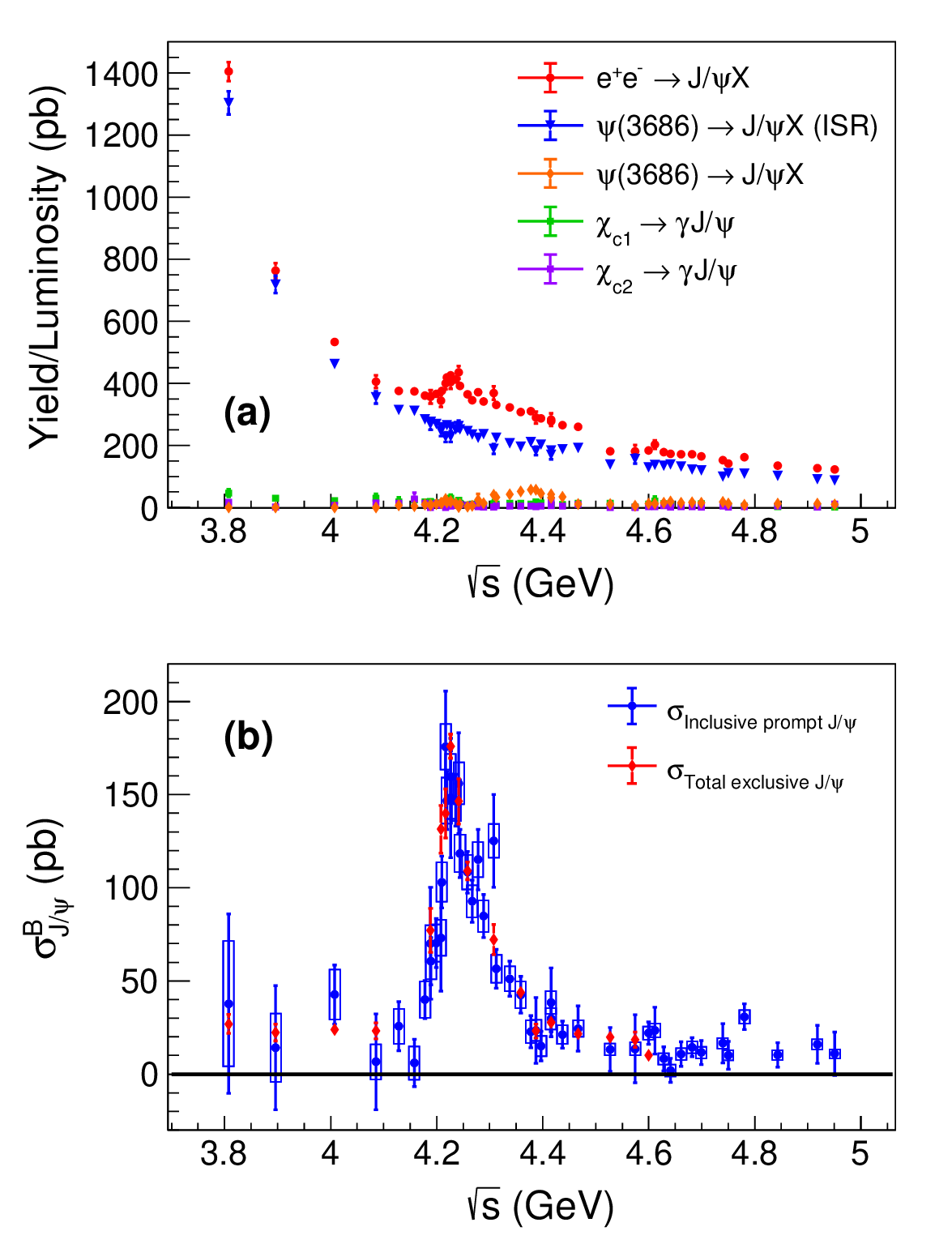}
\caption{(a)~The yields of $J/\psi$ mesons from different sources normalized to corresponding integrated luminosity. The sources are inclusive production of $J/\psi$ meson~(red dots), the ISR return to the $\psi(3686)$ resonance~(blue triangles), inclusive production of $\psi(3686)$ meson~(orange diamonds), and inclusive production of $\chi_{cJ}$~(\textit{J} = 1, 2) mesons~(green and violet boxes). (b)~Comparison of the inclusive Born cross section of prompt $J/\psi$ production~(blue dots) and the total Born cross section of the exclusive processes with a $J/\psi$ meson~(red diamonds). The uncertainties are total~(blue and red solid  lines) and systematic~(blue rectangles).}
\label{fig:Yield}
\end{center}
\end{figure}

\begin{figure}
\begin{center}
\includegraphics[width=0.5\textwidth]{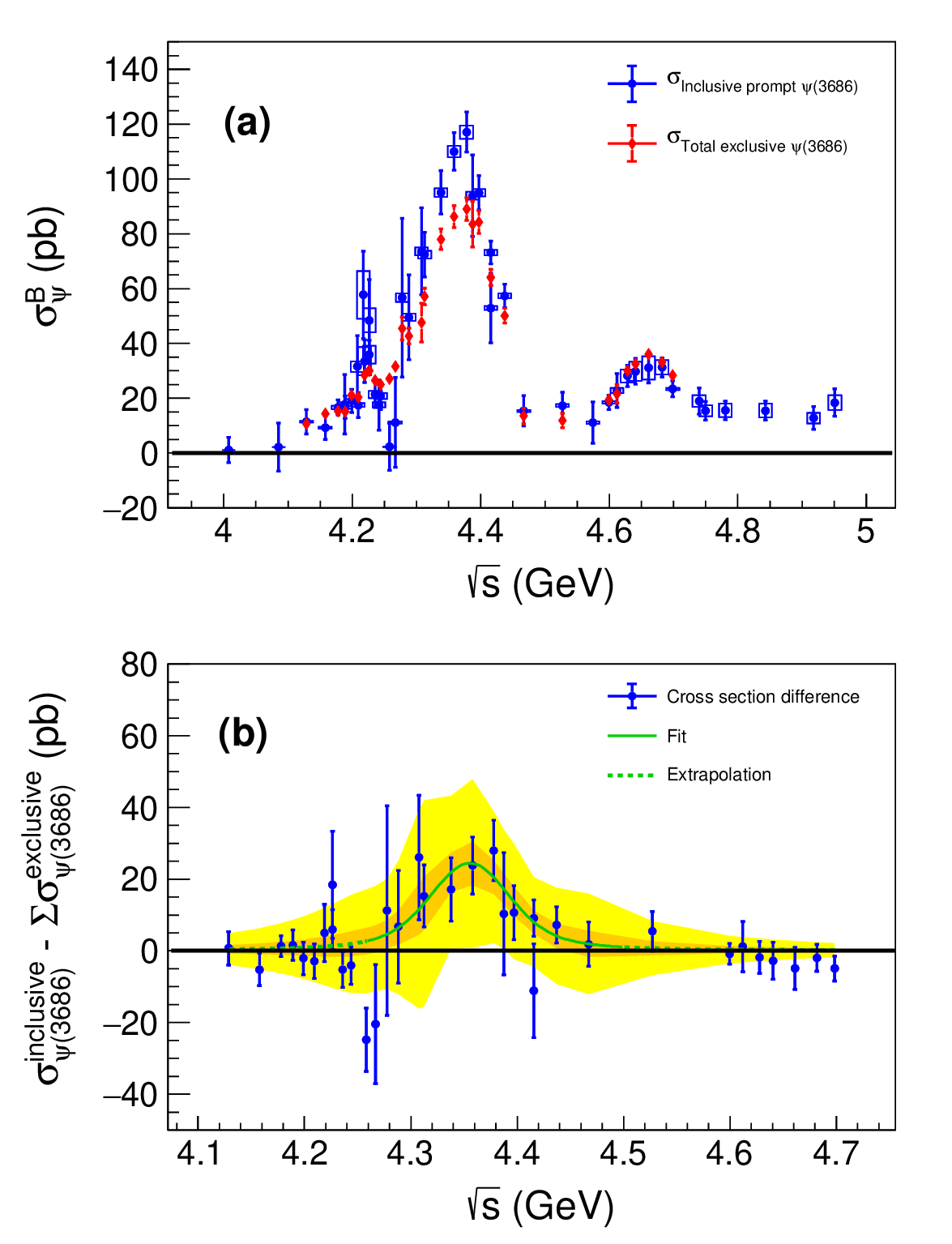}
\caption{(a)~Comparison of the inclusive Born cross section of prompt $\psi(3686)$ production~(blue dots) and the total Born cross section of the exclusive processes with $\psi(3686)$ meson~(red diamonds). The uncertainties are total~(blue and red solid  lines) and common systematic~(blue rectangles). (b)~The difference between the inclusive Born cross section of prompt $\psi(3686)$ production and the total Born cross section of the exclusive processes with a $\psi(3686)$ meson~(blue dots). The uncertainties are total. The fitting function and its $1\sigma$ and $3\sigma$ confidence intervals are shown by the green solid line and dark and light yellow areas, respectively. The extrapolation of the fitting function is shown by the green dashed line.}
\label{fig:TotalExcl_psiX}
\end{center}
\end{figure}

Applying Eqns.~(\ref{CS_JpsiX_prompt})$-$(\ref{CS_psiX_prompt}), we obtain the observed cross sections of the processes $e^{+}e^{-}\rightarrow J/\psi_{\textmd{prompt}}X$ and $e^{+}e^{-}\rightarrow \psi(3686)_{\textmd{prompt}}X$ at each energy point. Figure~\ref{fig:Yield}~(a) shows the yields of $J/\psi$ mesons from different sources normalized to the corresponding integrated luminosity, which are the terms of Eqn.~(\ref{CS_JpsiX_prompt}). The Born cross sections are derived from the observed ones by an iterative procedure given in Eqn.~(\ref{CS_Born})$-$(\ref{F_t}). The observed and Born cross sections of the processes $e^{+}e^{-}\rightarrow J/\psi_{\textmd{prompt}}X$ and $e^{+}e^{-}\rightarrow \psi(3686)_{\textmd{prompt}}X$, as well as the ISR correction and vacuum polarization factors, at each energy point are listed in Table~\ref{CS}. Figure~\ref{fig:Yield}~(b) shows a comparison of the inclusive Born cross section of prompt $J/\psi$ production and the total Born cross section of the available exclusive processes with $J/\psi$ meson such as $e^{+}e^{-}\rightarrow \pi^{+}\pi^{-}J/\psi$~\cite{Ablikim:2016qzw}, $e^{+}e^{-}\rightarrow \pi^{0}\pi^{0}J/\psi$~\cite{BESIII_Jpsi2pi0}, $e^{+}e^{-}\rightarrow \eta J/\psi$~\cite{BESIII_JpsiEtaPi0}, $e^{+}e^{-}\rightarrow \pi^{0}J/\psi$~\cite{BESIII_JpsiEtaPi0}, $e^{+}e^{-}\rightarrow K^{+}K^{-}J/\psi$~\cite{BESIII_Jpsi2K}, $e^{+}e^{-}\rightarrow K^{0}K^{0}J/\psi$~\cite{BESIII_Jpsi2K0}, $e^{+}e^{-}\rightarrow \eta' J/\psi$~\cite{BESIII_JpsiEtaPr}. The comparison indicate no evidence of hidden decays involving the $J/\psi$ meson, within the measurement uncertainties. Table~\ref{MES} shows a list of charmonium(-like) mesons from Ref.~\cite{pdg} decaying into $J/\psi$ or $\psi(3686)$. The wide skewed peaking structure is dominantly a superposition of contributions of the $\psi(4230)$ and $\psi(4360)$ mesons. The average value of the inclusive cross section of prompt $J/\psi$ production in the range from $\sqrt{s}=4.527$ to $4.951$~GeV is determined to be $14.0\pm1.7\pm3.1$~pb, assuming the absence of resonances decaying into the $J/\psi$ meson in that energy range.  
To obtain these values a constant is fitted to the Born cross sections only with statistical uncertainties at the fifteen highest energy points. The second quoted uncertainty is the averaged systematic uncertainty of the Born cross sections at the fifteen highest energy points. Figure~\ref{fig:TotalExcl_psiX}~(a) shows a comparison of the inclusive Born cross section of prompt $\psi(3686)$ production and the total Born cross section of the available exclusive processes with $\psi(3686)$ meson such as $e^{+}e^{-}\rightarrow \eta\psi(3686)$~\cite{BESIII_etaPsi} and  $e^{+}e^{-}\rightarrow \pi^{+}\pi^{-}\psi(3686)$~\cite{BESIII_Psi2pi}. The contribution of the $e^{+}e^{-}\rightarrow \pi^{+}\pi^{-}\psi(3686)$ channel is scaled by 1.5 times to take into account the isospin-symmetric $e^{+}e^{-}\rightarrow \pi^{0}\pi^{0}\psi(3686)$ channel as well. The three peaks correspond to the contributions of the $\psi(4230)$, $\psi(4360)$, and $\psi(4660)$ mesons. The average value of the inclusive cross section of prompt $\psi(3686)$ production in the range from $\sqrt{s}=4.843$ to $4.951$~GeV is determined to be $15.3\pm3.0$~pb, assuming the absence of resonances decaying into the $\psi(3686)$ meson in that energy range. 
 To obtain these values a constant is fitted to the Born cross sections only with independent uncertainties at the three highest energy points. The quoted total uncertainty is obtained by quadratically summing independent and common contributions. The common contribution is the averaged common uncertainty of the Born cross sections at the three highest energy points. Figure~\ref{fig:TotalExcl_psiX}~(b) shows the difference between the inclusive Born cross section of prompt $\psi(3686)$ production and the total Born cross section of the available exclusive processes with $\psi(3686)$ meson mentioned above. To evaluate the contribution of other possible decay channels of the $\psi(4360)$ meson to the $\psi(3686)X$ final states, a convolution of a Breit-Wigner curve with a Gaussian function is fitted to the difference in cross sections. The cross section of the process $e^{+}e^{-}\rightarrow \psi(3686)_{\textmd{prompt}}X$ with an unknown $X$ at a mass of the $\psi(4360)$ meson is about 23\% of the measured cross section for $e^+e^- \rightarrow \psi(3686)_{\textmd{prompt}} X$.

\begin{table*}[htbp]
\caption{The masses and widths of charmonium(-like) mesons with decays into $J/\psi$ or $\psi(3686)$~\cite{pdg,BESIII_Jpsi2K,BESIII_Jpsi2K0}. The notation~``$-$"  means that the corresponding decays have not yet seen.}
\begin{center}
\begin{tabular}{ccccc}\hline\hline
$c\bar{c}$ Meson & Mass~(MeV) & Width~(MeV) & Decays into $J/\psi$  & Decays into $\psi(3686)$ \\\hline
$\chi_{c1}(3872)$ & $3871.7\pm0.1$ & $\textcolor{white}{0}1.2\pm0.2\textcolor{white}{0}$ & $\pi^{+}\pi^{-}J/\psi, \omega J/\psi, \gamma J/\psi$ & $\gamma \psi(3686)$ \\\hline  
$Z_{c}(3900)$ & $3887.1\pm2.6$ & $28.4\pm2.6\textcolor{white}{0}$ & $\pi J/\psi$ & $-$ \\\hline 
$\chi_{c0}(3915)$ & $3921.7\pm1.8$ & $18.8\pm3.5\textcolor{white}{0}$ & $\omega J/\psi$ & $-$ \\\hline 
$\psi(4040)$ & $4039.0\pm1.0$ & $\textcolor{white}{.0}80\pm10\textcolor{white}{0.}$ & $\eta J/\psi$ & $-$ \\\hline
$X(4160)$ & $\textcolor{white}{.0}4153\pm23\textcolor{white}{.}$ & $\textcolor{white}{.}136\pm60\textcolor{white}{0.}$ & $\phi J/\psi$ & $-$ \\\hline 
$\psi(4230)$ & $4222.7\pm2.6$ & $49.0\pm8.0\textcolor{white}{0}$ & $\pi \pi J/\psi, K K J/\psi, \eta J/\psi$ & $\pi^{+}\pi^{-} \psi(3686)$ \\\hline 
$X(4350)$ & $4350.6\pm5.2$ & $\textcolor{white}{.0}13\pm18\textcolor{white}{.0}$ & $\phi J/\psi$ & $-$ \\\hline
$\psi(4360)$ & $4372.0\pm9.0$ & $\textcolor{white}{.}115\pm13\textcolor{white}{.0}$ & $\pi^{+}\pi^{-}J/\psi, \eta J/\psi$ & $\pi^{+}\pi^{-} \psi(3686)$ \\\hline 
$Y(4500)$ & $\textcolor{white}{.0}4485\pm28\textcolor{white}{.}$ & $\textcolor{white}{.}111\pm34\textcolor{white}{.0}$ & $K^{+}K^{-}J/\psi$ & $-$ \\\hline 
$\psi(4660)$ & $4630.0\pm6.0$ & $\textcolor{white}{.0}72\pm14\textcolor{white}{.0}$ & $-$ & $\pi^{+}\pi^{-} \psi(3686)$ \\\hline 
$Y(4710)$ & $\textcolor{white}{.0}4704\pm87\textcolor{white}{.}$ & $\textcolor{white}{.}183\pm149\textcolor{white}{.}$ & $K^{0}K^{0}J/\psi$ & $-$ \\\hline\hline 
\end{tabular}
\end{center}
\label{MES}
\end{table*}

\section{SUMMARY}

The inclusive cross sections of prompt $J/\psi$ and $\psi(3686)$ production are measured at center-of-mass energies from 3.808 to 4.951~GeV. The observed energy dependence of the cross section in the range below 4.843~GeV is mainly due to the production and decay of such resonances as $\psi(4230)$, $\psi(4360)$, and $\psi(4660)$. The measured inclusive cross section for the $J/\psi$ is in good agreement with the sum of the exclusive cross sections already measured. However, for the $\psi(3686)$, the contribution of unknown decay channels of the $\psi(4360)$ meson to the $\psi(3686)X$ final states is approximately 23\% of the measured inclusive cross section. The average values of the cross sections measured in the center-of-mass energy ranges from 4.527 to 4.951~GeV for $J/\psi$ and from 4.843 to 4.951~GeV for $\psi(3686)$, where the impact of known resonances is negligible, are $14.0\pm1.7\pm3.1$~pb and $15.3\pm3.0$~pb, respectively. For $J/\psi$, the first and the second uncertainties are statistical and systematic, respectively. For $\psi(3686)$, the uncertainty is total. The presented results can be useful for testing various phenomenological models of the strong interaction at low energies, especially, NRQCD.

\section*{Acknowledgement}

The BESIII Collaboration thanks the staff of BEPCII and the IHEP computing center for their strong support. This work is supported in part by National Key R\&D Program of China under Contracts Nos. 2020YFA0406300, 2020YFA0406400; National Natural Science Foundation of China (NSFC) under Contracts Nos. 11635010, 11735014, 11835012, 11935015, 11935016, 11935018, 11961141012, 12022510, 12025502, 12035009, 12035013, 12061131003, 12192260, 12192261, 12192262, 12192263, 12192264, 12192265, 12221005, 12225509, 12235017; the Chinese Academy of Sciences (CAS) Large-Scale Scientific Facility Program; the CAS Center for Excellence in Particle Physics (CCEPP); Joint Large-Scale Scientific Facility Funds of the NSFC and CAS under Contract No. U1832207; CAS Key Research Program of Frontier Sciences under Contracts Nos. QYZDJ-SSW-SLH003, QYZDJ-SSW-SLH040; 100 Talents Program of CAS; The Institute of Nuclear and Particle Physics (INPAC) and Shanghai Key Laboratory for Particle Physics and Cosmology; ERC under Contract No. 758462; European Union's Horizon 2020 research and innovation programme under Marie Sklodowska-Curie grant agreement under Contract No. 894790; German Research Foundation DFG under Contracts Nos. 455635585, Collaborative Research Center CRC 1044, FOR5327, GRK 2149; Istituto Nazionale di Fisica Nucleare, Italy; Ministry of Development of Turkey under Contract No. DPT2006K-120470; National Research Foundation of Korea under Contract No. NRF-2022R1A2C1092335; National Science and Technology fund of Mongolia; National Science Research and Innovation Fund (NSRF) via the Program Management Unit for Human Resources \& Institutional Development, Research and Innovation of Thailand under Contract No. B16F640076; Polish National Science Centre under Contract No. 2019/35/O/ST2/02907; The Swedish Research Council; U. S. Department of Energy under Contract No. DE-FG02-05ER41374.

\section*{APPENDIX}

Table~\ref{NoE} lists the numbers of events and ISR ratios necessary for calculating the observed cross sections, as a function of center-of-mass energy. Similarly, Table~\ref{EFF} gives the efficiencies and the observed cross sections obtained using the di-muon and di-electron decay modes.

\begin{table*}[htbp] \scriptsize   
\caption{The observed numbers of events, $N$, and the ISR coefficients, $\mathcal{R}$, used in formulae~(\ref{CS_JpsiX_prompt})-(\ref{CS_psiX_prompt}). The uncertainties are statistical for the $N$ and systematic for the $\mathcal{R}$.
}
\begin{center}
\begin{tabular}{cccccccccccc}\hline\hline
$\sqrt{s}$~(GeV) & $N_{J/\psi X}$ & $N_{\gamma_{\textmd{ISR}}J/\psi }$ & $\mathcal{R}_{\gamma_{\textmd{ISR}}J/\psi}~(\%)$  & $N^{\mu}_{\psi X}$ & $N^{\mu}_{\gamma_{\textmd{ISR}}\psi }$ & $\mathcal{R}^{\mu}_{\gamma_{\textmd{ISR}}\psi}~(\%)$ & $N^{e}_{\psi X}$ & $N^{e}_{\gamma_{\textmd{ISR}}\psi }$ & $\mathcal{R}^{e}_{\gamma_{\textmd{ISR}}\psi}~(\%)$ & $N_{\chi_{c1} X}$ & $N_{\chi_{c2} X}$ \\\hline
$3.808$ & $\textcolor{white}{0}3200\pm68\textcolor{white}{0}$ & $\textcolor{white}{0}2608\pm143$ & $1.94\pm0.08$ & $\textcolor{white}{00}56\pm8\textcolor{white}{00}$ & $\textcolor{white}{0}1272\pm37\textcolor{white}{0}$ & $5.1\pm0.4$ & $-$ & $-$ & $-$ & $\textcolor{white}{00}74\pm19\textcolor{white}{0}$ & $\textcolor{white}{00}23\pm12\textcolor{white}{0}$ \\\hline
$3.896$ & $\textcolor{white}{0}1815\pm57\textcolor{white}{0}$ & $\textcolor{white}{0}1627\pm133$ & $2.01\pm0.08$ & $\textcolor{white}{00}33\pm6\textcolor{white}{00}$ & $\textcolor{white}{00}727\pm28\textcolor{white}{0}$ & $5.5\pm0.4$ & $-$ & $-$ & $-$ & $\textcolor{white}{00}49\pm16\textcolor{white}{0}$ & $\textcolor{white}{000}2\pm9\textcolor{white}{00}$ \\\hline
$4.008$ & $11774\pm155$ & $15251\pm373$ & $2.58\pm0.10$ & $\textcolor{white}{0}270\pm19\textcolor{white}{0}$ & $\textcolor{white}{0}4203\pm69\textcolor{white}{0}$ & $6.5\pm0.5$ & $194\pm18 $ & $\textcolor{white}{0}2815\pm98\textcolor{white}{0}$ & $6.5\pm0.5$ & $\textcolor{white}{0}343\pm47\textcolor{white}{0}$ & $\textcolor{white}{00}96\pm29\textcolor{white}{0}$ \\\hline
$4.085$ & $\textcolor{white}{00}978\pm48\textcolor{white}{0}$ & $\textcolor{white}{0}1353\pm115$ & $2.20\pm0.09$ & $\textcolor{white}{00}21\pm6\textcolor{white}{00}$ & $\textcolor{white}{00}356\pm19\textcolor{white}{0}$ & $5.8\pm0.5$ & $\textcolor{white}{0}17\pm5\textcolor{white}{0}$ & $\textcolor{white}{00}237\pm27\textcolor{white}{0}$ & $6.1\pm0.5$ & $\textcolor{white}{00}49\pm24\textcolor{white}{0}$ & $\textcolor{white}{00}21\pm10\textcolor{white}{0}$ \\\hline
$4.129$ & $\textcolor{white}{0}6899\pm129$ & $10364\pm308$ & $2.73\pm0.11$ & $\textcolor{white}{0}203\pm17\textcolor{white}{0}$ & $\textcolor{white}{0}2278\pm50\textcolor{white}{0}$ & $6.8\pm0.5$ & $132\pm16$ & $\textcolor{white}{0}1531\pm68\textcolor{white}{0}$ & $6.9\pm0.6$ & $\textcolor{white}{0}226\pm190$ & $\textcolor{white}{0}141\pm101$ \\\hline
$4.158$ & $\textcolor{white}{0}6939\pm129$ & $10153\pm302$ & $2.28\pm0.09$ & $\textcolor{white}{0}178\pm17\textcolor{white}{0}$ & $\textcolor{white}{0}2340\pm51\textcolor{white}{0}$ & $6.2\pm0.5$ & $129\pm16$ & $\textcolor{white}{0}1544\pm67\textcolor{white}{0}$ & $6.3\pm0.5$ & $\textcolor{white}{0}318\pm90\textcolor{white}{0}$ & $\textcolor{white}{0}297\pm212$ \\\hline
$4.178$ & $51845\pm356$ & $78034\pm835$ & $2.21\pm0.09$ & $1582\pm53\textcolor{white}{0}$ & $16784\pm137$ & $5.9\pm0.5$ & $957\pm45$ & $11878\pm180$ & $5.9\pm0.5$ & $1821\pm115$ & $1168\pm133$ \\\hline
$4.189$ & $\textcolor{white}{00}704\pm41\textcolor{white}{0}$ & $\textcolor{white}{0}1035\pm97\textcolor{white}{0}$ & $2.28\pm0.09$ & $\textcolor{white}{00}20\pm6\textcolor{white}{00}$ & $\textcolor{white}{00}220\pm15\textcolor{white}{0}$ & $5.9\pm0.5$ & $\textcolor{white}{0}17\pm5\textcolor{white}{0}$ & $\textcolor{white}{00}163\pm22\textcolor{white}{0}$ & $6.2\pm0.5$ & $\textcolor{white}{00}26\pm11\textcolor{white}{0}$ & $\textcolor{white}{000}0\pm18\textcolor{white}{0}$ \\\hline
$4.189$ & $\textcolor{white}{0}8538\pm144$ & $11303\pm343$ & $2.17\pm0.09$ & $\textcolor{white}{0}250\pm21\textcolor{white}{0}$ & $\textcolor{white}{0}2661\pm55\textcolor{white}{0}$ & $6.1\pm0.5$ & $176\pm19$ & $\textcolor{white}{0}1884\pm72\textcolor{white}{0}$ & $6.1\pm0.5$ & $\textcolor{white}{0}277\pm50\textcolor{white}{0}$ & $\textcolor{white}{00}95\pm61\textcolor{white}{0}$ \\\hline
$4.199$ & $\textcolor{white}{0}8737\pm145$ & $12685\pm337$ & $2.10\pm0.08$ & $\textcolor{white}{0}271\pm23\textcolor{white}{0}$ & $\textcolor{white}{0}2629\pm55\textcolor{white}{0}$ & $6.0\pm0.5$ & $172\pm19$ & $\textcolor{white}{0}1743\pm71\textcolor{white}{0}$ & $6.3\pm0.5$ & $\textcolor{white}{0}252\pm51\textcolor{white}{0}$ & $\textcolor{white}{0}181\pm47\textcolor{white}{0}$ \\\hline
$4.208$ & $\textcolor{white}{00}856\pm47\textcolor{white}{0}$ & $\textcolor{white}{0}1140\pm107$ & $1.97\pm0.08$ & $\textcolor{white}{00}31\pm7\textcolor{white}{00}$ & $\textcolor{white}{00}258\pm17\textcolor{white}{0}$ & $5.6\pm0.4$ & $\textcolor{white}{0}23\pm6\textcolor{white}{0}$ & $\textcolor{white}{00}170\pm24\textcolor{white}{0}$ & $5.8\pm0.5$ & $\textcolor{white}{00}26\pm16\textcolor{white}{0}$ & $\textcolor{white}{00}13\pm15\textcolor{white}{0}$ \\\hline
$4.209$ & $\textcolor{white}{0}8771\pm146$ & $12945\pm335$ & $2.59\pm0.10$ & $\textcolor{white}{0}291\pm23\textcolor{white}{0}$ & $\textcolor{white}{0}2419\pm52\textcolor{white}{0}$ & $6.4\pm0.5$ & $147\pm20$ & $\textcolor{white}{0}1704\pm69\textcolor{white}{0}$ & $6.6\pm0.5$ & $\textcolor{white}{0}195\pm52\textcolor{white}{0}$ & $\textcolor{white}{00}92\pm42\textcolor{white}{0}$ \\\hline
$4.217$ & $\textcolor{white}{00}995\pm47\textcolor{white}{0}$ & $\textcolor{white}{0}1596\pm107$ & $2.02\pm0.08$ & $\textcolor{white}{00}43\pm8\textcolor{white}{00}$ & $\textcolor{white}{00}235\pm16\textcolor{white}{0}$ & $5.8\pm0.5$ & $\textcolor{white}{0}30\pm6\textcolor{white}{0}$ & $\textcolor{white}{00}168\pm23\textcolor{white}{0}$ & $6.0\pm0.5$ & $\textcolor{white}{00}13\pm13\textcolor{white}{0}$ & $\textcolor{white}{000}0\pm11\textcolor{white}{0}$ \\\hline
$4.219$ & $\textcolor{white}{0}9723\pm150$ & $12055\pm338$ & $2.63\pm0.11$ & $\textcolor{white}{0}332\pm26\textcolor{white}{0}$ & $\textcolor{white}{0}2488\pm53\textcolor{white}{0}$ & $6.7\pm0.5$ & $209\pm22$ & $\textcolor{white}{0}1646\pm68\textcolor{white}{0}$ & $6.8\pm0.5$ & $\textcolor{white}{0}230\pm53\textcolor{white}{0}$ & $\textcolor{white}{00}81\pm38\textcolor{white}{0}$ \\\hline
$4.226$ & $\textcolor{white}{00}815\pm44\textcolor{white}{0}$ & $\textcolor{white}{0}1076\pm97\textcolor{white}{0}$ & $2.07\pm0.08$ & $\textcolor{white}{00}30\pm7\textcolor{white}{00}$ & $\textcolor{white}{00}192\pm15\textcolor{white}{0}$ & $6.1\pm0.5$ & $\textcolor{white}{0}25\pm6\textcolor{white}{0}$ & $\textcolor{white}{00}148\pm21\textcolor{white}{0}$ & $5.8\pm0.5$ & $\textcolor{white}{00}43\pm15\textcolor{white}{0}$ & $\textcolor{white}{000}5\pm11\textcolor{white}{0}$ \\\hline
$4.226$ & $20277\pm214$ & $24586\pm472$ & $2.07\pm0.08$ & $\textcolor{white}{0}693\pm37\textcolor{white}{0}$ & $\textcolor{white}{0}5174\pm76\textcolor{white}{0}$ & $6.1\pm0.5$ & $454\pm31$ & $\textcolor{white}{0}3496\pm100$ & $5.8\pm0.5$ & $\textcolor{white}{0}424\pm76\textcolor{white}{0}$ & $\textcolor{white}{0}156\pm54\textcolor{white}{0}$ \\\hline
$4.236$ & $\textcolor{white}{0}9850\pm150$ & $12366\pm333$ & $2.14\pm0.09$ & $\textcolor{white}{0}290\pm27\textcolor{white}{0}$ & $\textcolor{white}{0}2486\pm53\textcolor{white}{0}$ & $6.1\pm0.5$ & $175\pm22$ & $\textcolor{white}{0}1703\pm69\textcolor{white}{0}$ & $6.0\pm0.5$ & $\textcolor{white}{0}264\pm54\textcolor{white}{0}$ & $\textcolor{white}{0}104\pm37\textcolor{white}{0}$ \\\hline
$4.242$ & $\textcolor{white}{0}1086\pm49\textcolor{white}{0}$ & $\textcolor{white}{0}1044\pm107$ & $2.03\pm0.08$ & $\textcolor{white}{00}19\pm6\textcolor{white}{00}$ & $\textcolor{white}{00}278\pm18\textcolor{white}{0}$ & $5.9\pm0.5$ & $\textcolor{white}{0}31\pm7\textcolor{white}{0}$ & $\textcolor{white}{00}170\pm23\textcolor{white}{0}$ & $5.8\pm0.5$ & $\textcolor{white}{00}40\pm17\textcolor{white}{0}$ & $\textcolor{white}{00}15\pm12\textcolor{white}{0}$ \\\hline
$4.244$ & $\textcolor{white}{0}9460\pm151$ & $11107\pm340$ & $2.06\pm0.08$ & $\textcolor{white}{0}275\pm29\textcolor{white}{0}$ & $\textcolor{white}{0}2526\pm53\textcolor{white}{0}$ & $6.1\pm0.5$ & $182\pm25$ & $\textcolor{white}{0}1739\pm68\textcolor{white}{0}$ & $6.1\pm0.5$ & $\textcolor{white}{0}224\pm53\textcolor{white}{0}$ & $\textcolor{white}{0}134\pm37\textcolor{white}{0}$ \\\hline
$4.258$ & $13741\pm185$ & $18552\pm418$ & $2.62\pm0.10$ & $\textcolor{white}{0}309\pm90\textcolor{white}{0}$ & $\textcolor{white}{0}3760\pm65\textcolor{white}{0}$ & $6.9\pm0.6$ & $149\pm78$ & $\textcolor{white}{0}2479\pm85\textcolor{white}{0}$ & $6.7\pm0.5$ & $\textcolor{white}{0}234\pm64\textcolor{white}{0}$ & $\textcolor{white}{0}153\pm47\textcolor{white}{0}$ \\\hline
$4.267$ & $\textcolor{white}{0}8289\pm145$ & $11967\pm327$ & $2.04\pm0.08$ & $\textcolor{white}{0}193\pm109$ & $\textcolor{white}{0}2330\pm51\textcolor{white}{0}$ & $5.8\pm0.5$ & $133\pm96$ & $\textcolor{white}{0}1556\pm66\textcolor{white}{0}$ & $5.9\pm0.5$ & $\textcolor{white}{0}180\pm47\textcolor{white}{0}$ & $\textcolor{white}{0}113\pm38\textcolor{white}{0}$ \\\hline
$4.278$ & $\textcolor{white}{0}2936\pm86\textcolor{white}{0}$ & $\textcolor{white}{0}3357\pm187$ & $2.59\pm0.10$ & $\textcolor{white}{0}133\pm60\textcolor{white}{0}$ & $\textcolor{white}{00}717\pm28\textcolor{white}{0}$ & $6.7\pm0.5$ & $114\pm54$ & $\textcolor{white}{00}552\pm38\textcolor{white}{0}$ & $6.8\pm0.5$ & $\textcolor{white}{00}65\pm60\textcolor{white}{0}$ & $\textcolor{white}{00}14\pm21\textcolor{white}{0}$ \\\hline
$4.288$ & $\textcolor{white}{0}7724\pm142$ & $\textcolor{white}{0}9769\pm329$ & $2.72\pm0.11$ & $\textcolor{white}{0}273\pm99\textcolor{white}{0}$ & $\textcolor{white}{0}2118\pm48\textcolor{white}{0}$ & $6.7\pm0.5$ & $322\pm69$ & $\textcolor{white}{0}1427\pm63\textcolor{white}{0}$ & $6.8\pm0.5$ & $\textcolor{white}{0}136\pm51\textcolor{white}{0}$ & $\textcolor{white}{00}28\pm34\textcolor{white}{0}$ \\\hline
$4.308$ & $\textcolor{white}{00}750\pm42\textcolor{white}{0}$ & $\textcolor{white}{00}970\pm94\textcolor{white}{0}$ & $2.17\pm0.09$ & $\textcolor{white}{00}46\pm8\textcolor{white}{00}$ & $\textcolor{white}{00}159\pm13\textcolor{white}{0}$ & $6.2\pm0.5$ & $\textcolor{white}{0}27\pm6\textcolor{white}{0}$ & $\textcolor{white}{00}118\pm19\textcolor{white}{0}$ & $6.2\pm0.5$ & $\textcolor{white}{00}17\pm14\textcolor{white}{0}$ & $\textcolor{white}{000}1\pm10\textcolor{white}{0}$ \\\hline
$4.313$ & $\textcolor{white}{0}7471\pm142$ & $10429\pm309$ & $2.65\pm0.11$ & $\textcolor{white}{0}440\pm35\textcolor{white}{0}$ & $\textcolor{white}{0}2013\pm47\textcolor{white}{0}$ & $6.8\pm0.5$ & $347\pm29$ & $\textcolor{white}{0}1480\pm62\textcolor{white}{0}$ & $6.9\pm0.6$ & $\textcolor{white}{0}168\pm50\textcolor{white}{0}$ & $\textcolor{white}{00}74\pm34\textcolor{white}{0}$ \\\hline
$4.338$ & $\textcolor{white}{0}7331\pm141$ & $10531\pm305$ & $2.33\pm0.09$ & $\textcolor{white}{0}518\pm30\textcolor{white}{0}$ & $\textcolor{white}{0}1895\pm46\textcolor{white}{0}$ & $6.3\pm0.5$ & $399\pm25$ & $\textcolor{white}{0}1246\pm61\textcolor{white}{0}$ & $6.4\pm0.5$ & $\textcolor{white}{0}206\pm47\textcolor{white}{0}$ & $\textcolor{white}{00}59\pm32\textcolor{white}{0}$ \\\hline
$4.358$ & $\textcolor{white}{0}7518\pm144$ & $10493\pm319$ & $2.04\pm0.08$ & $\textcolor{white}{0}670\pm33\textcolor{white}{0}$ & $\textcolor{white}{0}1991\pm47\textcolor{white}{0}$ & $6.0\pm0.5$ & $495\pm26$ & $\textcolor{white}{0}1284\pm63\textcolor{white}{0}$ & $6.2\pm0.5$ & $\textcolor{white}{0}230\pm54\textcolor{white}{0}$ & $\textcolor{white}{00}54\pm32\textcolor{white}{0}$ \\\hline
$4.378$ & $\textcolor{white}{0}7275\pm144$ & $10347\pm307$ & $2.40\pm0.10$ & $\textcolor{white}{0}687\pm33\textcolor{white}{0}$ & $\textcolor{white}{0}1988\pm47\textcolor{white}{0}$ & $6.2\pm0.5$ & $497\pm26$ & $\textcolor{white}{0}1296\pm60\textcolor{white}{0}$ & $6.3\pm0.5$ & $\textcolor{white}{0}171\pm49\textcolor{white}{0}$ & $\textcolor{white}{00}51\pm32\textcolor{white}{0}$ \\\hline
$4.387$ & $\textcolor{white}{00}719\pm45\textcolor{white}{0}$ & $\textcolor{white}{00}861\pm98\textcolor{white}{0}$ & $2.01\pm0.08$ & $\textcolor{white}{00}73\pm11\textcolor{white}{0}$ & $\textcolor{white}{00}190\pm14\textcolor{white}{0}$ & $5.7\pm0.5$ & $\textcolor{white}{0}37\pm8\textcolor{white}{0}$ & $\textcolor{white}{00}124\pm19\textcolor{white}{0}$ & $5.9\pm0.5$ & $\textcolor{white}{00}31\pm14\textcolor{white}{0}$ & $\textcolor{white}{000}5\pm9\textcolor{white}{00}$ \\\hline
$4.397$ & $\textcolor{white}{0}6527\pm140$ & $\textcolor{white}{0}9033\pm299$ & $2.37\pm0.09$ & $\textcolor{white}{0}557\pm31\textcolor{white}{0}$ & $\textcolor{white}{0}1841\pm45\textcolor{white}{0}$ & $6.5\pm0.5$ & $468\pm26$ & $\textcolor{white}{0}1228\pm58\textcolor{white}{0}$ & $6.4\pm0.5$ & $\textcolor{white}{0}247\pm51\textcolor{white}{0}$ & $\textcolor{white}{00}55\pm31\textcolor{white}{0}$ \\\hline
$4.416$ & $\textcolor{white}{00}592\pm41\textcolor{white}{0}$ & $\textcolor{white}{00}691\pm89\textcolor{white}{0}$ & $2.07\pm0.08$ & $\textcolor{white}{00}37\pm9\textcolor{white}{00}$ & $\textcolor{white}{00}149\pm13\textcolor{white}{0}$ & $5.9\pm0.5$ & $\textcolor{white}{0}24\pm7\textcolor{white}{0}$ & $\textcolor{white}{000}87\pm17\textcolor{white}{0}$ & $6.0\pm0.5$ & $\textcolor{white}{00}12\pm13\textcolor{white}{0}$ & $\textcolor{white}{00}29\pm10\textcolor{white}{0}$ \\\hline
$4.416$ & $13065\pm197$ & $19384\pm416$ & $2.07\pm0.08$ & $1080\pm43\textcolor{white}{0}$ & $\textcolor{white}{0}3584\pm64\textcolor{white}{0}$ & $5.9\pm0.5$ & $717\pm33$ & $\textcolor{white}{0}2405\pm83\textcolor{white}{0}$ & $6.0\pm0.5$ & $\textcolor{white}{0}268\pm65\textcolor{white}{0}$ & $\textcolor{white}{0}141\pm43\textcolor{white}{0}$ \\\hline
$4.437$ & $\textcolor{white}{0}6796\pm147$ & $10207\pm322$ & $2.21\pm0.09$ & $\textcolor{white}{0}497\pm31\textcolor{white}{0}$ & $\textcolor{white}{0}1932\pm46\textcolor{white}{0}$ & $6.3\pm0.5$ & $359\pm24$ & $\textcolor{white}{0}1342\pm60\textcolor{white}{0}$ & $6.3\pm0.5$ & $\textcolor{white}{0}182\pm53\textcolor{white}{0}$ & $\textcolor{white}{00}72\pm33\textcolor{white}{0}$ \\\hline
$4.467$ & $\textcolor{white}{0}1285\pm64\textcolor{white}{0}$ & $\textcolor{white}{0}1685\pm135$ & $2.02\pm0.08$ & $\textcolor{white}{00}47\pm11\textcolor{white}{0}$ & $\textcolor{white}{00}392\pm21\textcolor{white}{0}$ & $5.9\pm0.5$ & $\textcolor{white}{0}30\pm8\textcolor{white}{0}$ & $\textcolor{white}{00}253\pm26\textcolor{white}{0}$ & $6.1\pm0.5$ & $\textcolor{white}{00}49\pm19\textcolor{white}{0}$ & $\textcolor{white}{00}27\pm13\textcolor{white}{0}$ \\\hline
$4.527$ & $\textcolor{white}{00}922\pm63\textcolor{white}{0}$ & $\textcolor{white}{0}2066\pm135$ & $2.17\pm0.09$ & $\textcolor{white}{00}40\pm10\textcolor{white}{0}$ & $\textcolor{white}{00}287\pm18\textcolor{white}{0}$ & $6.2\pm0.5$ & $\textcolor{white}{0}36\pm8\textcolor{white}{0}$ & $\textcolor{white}{00}221\pm24\textcolor{white}{0}$ & $6.0\pm0.5$ & $\textcolor{white}{00}46\pm21\textcolor{white}{0}$ & $\textcolor{white}{000}0\pm12\textcolor{white}{0}$ \\\hline
$4.575$ & $\textcolor{white}{00}394\pm42\textcolor{white}{0}$ & $\textcolor{white}{00}521\pm89\textcolor{white}{0}$ & $2.02\pm0.08$ & $\textcolor{white}{00}14\pm7\textcolor{white}{00}$ & $\textcolor{white}{00}139\pm12\textcolor{white}{0}$ & $6.0\pm0.5$ & $\textcolor{white}{0}11\pm5\textcolor{white}{0}$ & $\textcolor{white}{000}88\pm16\textcolor{white}{0}$ & $5.9\pm0.5$ & $\textcolor{white}{000}2\pm10\textcolor{white}{0}$ & $\textcolor{white}{000}0\pm11\textcolor{white}{0}$ \\\hline
$4.600$ & $\textcolor{white}{0}4846\pm145$ & $\textcolor{white}{0}9343\pm306$ & $2.02\pm0.08$ & $\textcolor{white}{0}232\pm26\textcolor{white}{0}$ & $\textcolor{white}{0}1392\pm40\textcolor{white}{0}$ & $5.9\pm0.5$ & $157\pm17$ & $\textcolor{white}{00}996\pm52\textcolor{white}{0}$ & $6.0\pm0.5$ & $\textcolor{white}{0}195\pm48\textcolor{white}{0}$ & $\textcolor{white}{00}39\pm30\textcolor{white}{0}$ \\\hline
$4.612$ & $\textcolor{white}{00}937\pm61\textcolor{white}{0}$ & $\textcolor{white}{0}1543\pm124$ & $2.12\pm0.08$ & $\textcolor{white}{00}40\pm11\textcolor{white}{0}$ & $\textcolor{white}{00}251\pm17\textcolor{white}{0}$ & $6.2\pm0.5$ & $\textcolor{white}{0}35\pm7\textcolor{white}{0}$ & $\textcolor{white}{00}179\pm22\textcolor{white}{0}$ & $6.3\pm0.5$ & $\textcolor{white}{00}65\pm48\textcolor{white}{0}$ & $\textcolor{white}{000}7\pm16\textcolor{white}{0}$ \\\hline
$4.628$ & $\textcolor{white}{0}4150\pm138$ & $\textcolor{white}{0}7498\pm281$ & $2.22\pm0.09$ & $\textcolor{white}{0}223\pm24\textcolor{white}{0}$ & $\textcolor{white}{0}1227\pm37\textcolor{white}{0}$ & $6.2\pm0.5$ & $195\pm18$ & $\textcolor{white}{00}883\pm48\textcolor{white}{0}$ & $6.5\pm0.5$ & $\textcolor{white}{0}202\pm48\textcolor{white}{0}$ & $\textcolor{white}{00}67\pm31\textcolor{white}{0}$ \\\hline
$4.641$ & $\textcolor{white}{0}4350\pm144$ & $\textcolor{white}{0}8615\pm283$ & $2.78\pm0.11$ & $\textcolor{white}{0}290\pm26\textcolor{white}{0}$ & $\textcolor{white}{0}1346\pm39\textcolor{white}{0}$ & $6.7\pm0.5$ & $180\pm17$ & $\textcolor{white}{00}895\pm48\textcolor{white}{0}$ & $6.8\pm0.5$ & $\textcolor{white}{0}183\pm49\textcolor{white}{0}$ & $\textcolor{white}{00}20\pm31\textcolor{white}{0}$ \\\hline
$4.661$ & $\textcolor{white}{0}4037\pm140$ & $\textcolor{white}{0}7176\pm277$ & $2.24\pm0.09$ & $\textcolor{white}{0}233\pm25\textcolor{white}{0}$ & $\textcolor{white}{0}1222\pm37\textcolor{white}{0}$ & $6.3\pm0.5$ & $197\pm17$ & $\textcolor{white}{00}872\pm48\textcolor{white}{0}$ & $6.3\pm0.5$ & $\textcolor{white}{00}92\pm33\textcolor{white}{0}$ & $\textcolor{white}{00}47\pm30\textcolor{white}{0}$ \\\hline
$4.682$ & $12697\pm315$ & $21563\pm511$ & $2.08\pm0.08$ & $\textcolor{white}{0}733\pm45\textcolor{white}{0}$ & $\textcolor{white}{0}3565\pm64\textcolor{white}{0}$ & $6.2\pm0.5$ & $648\pm32$ & $\textcolor{white}{0}2486\pm83\textcolor{white}{0}$ & $6.2\pm0.5$ & $\textcolor{white}{0}557\pm84\textcolor{white}{0}$ & $\textcolor{white}{0}152\pm55\textcolor{white}{0}$ \\\hline
$4.699$ & $\textcolor{white}{0}3918\pm142$ & $\textcolor{white}{0}7599\pm276$ & $2.10\pm0.08$ & $224\pm25$ & $1119\pm36$ & $6.0\pm0.5$ & $157\pm16$ & $\textcolor{white}{00}848\pm47\textcolor{white}{0}$ & $6.1\pm0.5$ & $\textcolor{white}{0}205\pm35\textcolor{white}{0}$ & $\textcolor{white}{00}24\pm30\textcolor{white}{0}$ \\\hline
$4.740$ & $\textcolor{white}{0}1118\pm79\textcolor{white}{0}$ & $\textcolor{white}{0}2277\pm152$ & $2.16\pm0.09$ & $\textcolor{white}{00}76\pm15\textcolor{white}{0}$ & $\textcolor{white}{00}290\pm19\textcolor{white}{0}$ & $6.1\pm0.5$ & $\textcolor{white}{0}38\pm9\textcolor{white}{0}$ & $\textcolor{white}{00}229\pm24\textcolor{white}{0}$ & $6.3\pm0.5$ & $\textcolor{white}{00}50\pm20\textcolor{white}{0}$ & $\textcolor{white}{00}16\pm18\textcolor{white}{0}$ \\\hline
$4.750$ & $\textcolor{white}{0}2342\pm118$ & $\textcolor{white}{0}5148\pm222$ & $2.20\pm0.09$ & $\textcolor{white}{0}131\pm21\textcolor{white}{0}$ & $\textcolor{white}{00}727\pm29\textcolor{white}{0}$ & $6.2\pm0.5$ & $\textcolor{white}{0}88\pm13$ & $\textcolor{white}{00}571\pm37\textcolor{white}{0}$ & $6.1\pm0.5$ & $\textcolor{white}{00}59\pm28\textcolor{white}{0}$ & $\textcolor{white}{00}13\pm25\textcolor{white}{0}$ \\\hline
$4.781$ & $\textcolor{white}{0}3705\pm141$ & $\textcolor{white}{0}6687\pm263$ & $2.19\pm0.09$ & $\textcolor{white}{0}145\pm25\textcolor{white}{0}$ & $\textcolor{white}{0}1000\pm34\textcolor{white}{0}$ & $6.0\pm0.5$ & $143\pm15$ & $\textcolor{white}{00}728\pm43\textcolor{white}{0}$ & $6.2\pm0.5$ & $\textcolor{white}{00}63\pm34\textcolor{white}{0}$ & $\textcolor{white}{00}36\pm30\textcolor{white}{0}$ \\\hline
$4.843$ & $\textcolor{white}{0}3144\pm144$ & $\textcolor{white}{0}5852\pm266$ & $2.25\pm0.09$ & $\textcolor{white}{0}183\pm26\textcolor{white}{0}$ & $\textcolor{white}{00}956\pm34\textcolor{white}{0}$ & $6.1\pm0.5$ & $123\pm15$ & $\textcolor{white}{00}708\pm43\textcolor{white}{0}$ & $6.3\pm0.5$ & $\textcolor{white}{00}47\pm35\textcolor{white}{0}$ & $\textcolor{white}{00}65\pm32\textcolor{white}{0}$ \\\hline
$4.918$ & $\textcolor{white}{0}1152\pm91\textcolor{white}{0}$ & $\textcolor{white}{0}1920\pm164$ & $2.04\pm0.08$ & $\textcolor{white}{00}69\pm16\textcolor{white}{0}$ & $\textcolor{white}{00}337\pm20\textcolor{white}{0}$ & $5.9\pm0.5$ & $\textcolor{white}{0}39\pm9\textcolor{white}{0}$ & $\textcolor{white}{00}286\pm26\textcolor{white}{0}$ & $6.2\pm0.5$ & $\textcolor{white}{000}6\pm23\textcolor{white}{0}$ & $\textcolor{white}{00}12\pm21\textcolor{white}{0}$ \\\hline
$4.951$ & $\textcolor{white}{00}874\pm81\textcolor{white}{0}$ & $\textcolor{white}{0}1897\pm149$ & $2.41\pm0.10$ & $\textcolor{white}{00}45\pm14\textcolor{white}{0}$ & $\textcolor{white}{00}241\pm17\textcolor{white}{0}$ & $6.3\pm0.5$ & $\textcolor{white}{0}46\pm9\textcolor{white}{0}$ & $\textcolor{white}{00}173\pm21\textcolor{white}{0}$ & $6.5\pm0.5$ & $\textcolor{white}{00}11\pm20\textcolor{white}{0}$ & $\textcolor{white}{00}44\pm19\textcolor{white}{0}$ \\\hline\hline
\end{tabular}
\end{center}
\label{NoE}
\end{table*}

\begin{table*}[htbp]
\caption{The various reconstruction efficiencies, $\epsilon$, used in formulae~(\ref{CS_JpsiX_prompt})-(\ref{CS_psiX_prompt}), and the observed cross sections obtained using the di-muon, $\sigma^{O}_{\psi, \mu}$, and di-electron, $\sigma^{O}_{\psi, e}$, decay modes. For the $\epsilon$, the uncertainties are systematic. For the $\sigma^{O}_{\psi, l}$, the first and the second uncertainties are statistical and independent systematic, respectively.}
\begin{center}
\begin{tabular}{ccccccccc}\hline\hline
$\sqrt{s}$~(GeV) & $\epsilon_{J/\psi X}~(\%)$ & $\epsilon^{\mu}_{\psi X}~(\%)$  & $\epsilon^{e}_{\psi X}~(\%)$  & $\epsilon_{\gamma_{\textmd{ISR}} \psi}~(\%)$ & $\epsilon_{\chi_{c1} X}~(\%)$ & $\epsilon_{\chi_{c2} X}~(\%)$ & $\sigma^{O}_{\psi, \mu}$~(pb) & $\sigma^{O}_{\psi, e}$~(pb) \\\hline
$3.808$ & $74.4\pm1.5$ & $52.8\pm2.1$ & $-$ & $52.6\pm2.1$ & $52.4\pm1.2$ & $44.4\pm1.0$ & $-$ & $-$ \\\hline
$3.896$ & $74.4\pm1.5$ & $52.7\pm2.1$ & $-$ & $52.3\pm2.1$ & $52.9\pm1.2$ & $44.4\pm1.0$ & $-$ & $-$ \\\hline
$4.008$ & $74.2\pm1.5$ & $51.5\pm2.1$ & $39.2\pm1.6$ & $51.3\pm2.1$ & $52.8\pm1.2$ & $44.7\pm1.0$ & $\textcolor{white}{00}-0.3\pm3.8\pm4.2$ & $\textcolor{white}{-00}2.7\pm5.0\pm3.9$ \\\hline
$4.085$ & $74.2\pm1.5$ & $52.7\pm2.1$ & $39.0\pm1.6$ & $51.6\pm2.1$ & $52.3\pm1.2$ & $44.1\pm1.0$ & $\textcolor{white}{-00}0.1\pm9.8\pm2.9$ & $\textcolor{white}{-.000}5\pm13\textcolor{white}{.}\pm3.3$ \\\hline
$4.129$ & $73.5\pm1.5$ & $50.9\pm2.0$ & $38.1\pm1.5$ & $49.0\pm2.0$ & $52.3\pm1.2$ & $44.3\pm1.0$ & $\textcolor{white}{-0}11.6\pm4.2\pm3.0$ & $\textcolor{white}{-00}8.4\pm5.2\pm4.1$ \\\hline
$4.158$ & $73.6\pm1.5$ & $51.2\pm2.0$ & $38.5\pm1.5$ & $50.0\pm2.0$ & $52.3\pm1.1$ & $43.9\pm1.0$ & $\textcolor{white}{-00}7.7\pm4.1\pm2.7$ & $\textcolor{white}{-00}9.6\pm5.0\pm4.3$ \\\hline
$4.178$ & $73.1\pm1.5$ & $52.1\pm2.1$ & $38.6\pm1.5$ & $50.4\pm2.0$ & $52.0\pm1.1$ & $43.7\pm1.0$ & $\textcolor{white}{-0}17.0\pm1.6\pm2.5$ & $\textcolor{white}{-0}10.0\pm1.8\pm4.3$ \\\hline
$4.189$ & $73.9\pm1.5$ & $53.0\pm2.1$ & $39.3\pm1.6$ & $51.2\pm2.0$ & $52.3\pm1.2$ & $44.0\pm1.0$ & $\textcolor{white}{-.00}14\pm12\textcolor{white}{.}\pm3.1$ & $\textcolor{white}{-.00}19\pm15\textcolor{white}{.}\pm2.5$ \\\hline
$4.189$ & $73.5\pm1.5$ & $51.8\pm2.1$ & $38.5\pm1.5$ & $49.8\pm2.0$ & $52.0\pm1.1$ & $43.8\pm1.0$ & $\textcolor{white}{-0}15.6\pm3.8\pm3.3$ & $\textcolor{white}{-0}14.6\pm4.6\pm2.3$ \\\hline
$4.199$ & $73.7\pm1.5$ & $52.3\pm2.1$ & $38.9\pm1.6$ & $50.5\pm2.0$ & $52.1\pm1.1$ & $43.7\pm1.0$ & $\textcolor{white}{-0}20.0\pm4.0\pm3.7$ & $\textcolor{white}{-0}14.6\pm4.7\pm2.2$ \\\hline
$4.208$ & $73.9\pm1.5$ & $53.4\pm2.1$ & $40.0\pm1.6$ & $51.7\pm2.1$ & $52.2\pm1.1$ & $43.7\pm1.0$ & $\textcolor{white}{-.00}27\pm12\textcolor{white}{.}\pm4.5$ & $\textcolor{white}{-.00}29\pm15\textcolor{white}{.}\pm2.2$ \\\hline
$4.209$ & $72.8\pm1.5$ & $51.2\pm2.0$ & $38.1\pm1.5$ & $49.2\pm2.0$ & $51.7\pm1.1$ & $43.6\pm1.0$ & $\textcolor{white}{-0}24.8\pm4.3\pm4.4$ & $\textcolor{white}{-00}8.4\pm5.0\pm2.2$ \\\hline
$4.217$ & $73.8\pm1.5$ & $53.2\pm2.1$ & $40.0\pm1.6$ & $51.5\pm2.1$ & $52.1\pm1.1$ & $44.0\pm1.0$ & $\textcolor{white}{-.00}48\pm13\textcolor{white}{.}\pm7.5$ & $\textcolor{white}{-.00}44\pm15\textcolor{white}{.}\pm2.8$ \\\hline
$4.219$ & $73.2\pm1.5$ & $51.4\pm2.1$ & $38.6\pm1.5$ & $49.5\pm2.0$ & $51.7\pm1.1$ & $43.8\pm1.0$ & $\textcolor{white}{-0}30.3\pm4.8\pm5.2$ & $\textcolor{white}{-0}23.7\pm5.5\pm2.5$ \\\hline
$4.226$ & $73.8\pm1.5$ & $53.0\pm2.1$ & $39.7\pm1.6$ & $51.2\pm2.0$ & $52.2\pm1.1$ & $43.8\pm1.0$ & $\textcolor{white}{-.00}37\pm15\textcolor{white}{.}\pm5.9$ & $\textcolor{white}{-.00}45\pm18\textcolor{white}{.}\pm2.9$ \\\hline
$4.226$ & $73.8\pm1.5$ & $53.0\pm2.1$ & $39.7\pm1.6$ & $51.2\pm2.0$ & $52.2\pm1.1$ & $43.8\pm1.0$ & $\textcolor{white}{-0}32.7\pm3.2\pm5.4$ & $\textcolor{white}{-0}29.0\pm3.6\pm2.3$ \\\hline
$4.236$ & $73.5\pm1.5$ & $52.7\pm2.1$ & $39.2\pm1.6$ & $50.8\pm2.0$ & $52.1\pm1.1$ & $43.8\pm1.0$ & $\textcolor{white}{-0}24.1\pm4.7\pm4.2$ & $\textcolor{white}{-0}16.8\pm5.2\pm2.1$ \\\hline
$4.242$ & $73.5\pm1.5$ & $53.4\pm2.1$ & $40.0\pm1.6$ & $51.5\pm2.1$ & $52.0\pm1.1$ & $43.8\pm1.0$ & $\textcolor{white}{-00}3.6\pm9.9\pm2.2$ & $\textcolor{white}{-.00}45\pm15\textcolor{white}{.}\pm2.8$ \\\hline
$4.244$ & $73.4\pm1.5$ & $52.5\pm2.1$ & $39.1\pm1.6$ & $50.6\pm2.0$ & $51.9\pm1.1$ & $43.6\pm1.0$ & $\textcolor{white}{-0}20.7\pm5.0\pm3.8$ & $\textcolor{white}{-0}17.5\pm5.9\pm2.1$ \\\hline
$4.258$ & $73.6\pm1.5$ & $52.9\pm2.1$ & $39.8\pm1.6$ & $50.1\pm2.0$ & $52.0\pm1.1$ & $44.0\pm1.0$ & $\textcolor{white}{-.000}5\pm10\textcolor{white}{.}\pm2.4$ & $\textcolor{white}{.000}-2\pm12\textcolor{white}{.}\pm1.9$ \\\hline
$4.267$ & $73.6\pm1.5$ & $52.7\pm2.1$ & $39.3\pm1.6$ & $50.7\pm2.0$ & $51.9\pm1.1$ & $43.8\pm1.0$ & $\textcolor{white}{-.00}10\pm19\textcolor{white}{.}\pm2.4$ & $\textcolor{white}{-.000}9\pm22\textcolor{white}{.}\pm1.8$ \\\hline
$4.278$ & $73.2\pm1.5$ & $51.8\pm2.1$ & $38.6\pm1.5$ & $49.5\pm2.0$ & $51.6\pm1.1$ & $43.7\pm1.0$ & $\textcolor{white}{-.00}45\pm32\textcolor{white}{.}\pm7.1$ & $\textcolor{white}{-.00}55\pm39\textcolor{white}{.}\pm3.4$ \\\hline
$4.288$ & $72.8\pm1.5$ & $50.4\pm2.0$ & $38.0\pm1.5$ & $48.4\pm1.9$ & $51.8\pm1.1$ & $43.7\pm1.0$ & $\textcolor{white}{-.00}25\pm19\textcolor{white}{.}\pm4.4$ & $\textcolor{white}{-.00}57\pm18\textcolor{white}{.}\pm3.4$ \\\hline
$4.308$ & $73.5\pm1.5$ & $53.1\pm2.1$ & $39.8\pm1.6$ & $50.8\pm2.0$ & $51.8\pm1.1$ & $43.9\pm1.0$ & $\textcolor{white}{-.00}73\pm17\textcolor{white}{.}\pm11\textcolor{white}{.}$ & $\textcolor{white}{-.00}53\pm17\textcolor{white}{.}\pm3.0$ \\\hline
$4.313$ & $72.8\pm1.5$ & $50.8\pm2.0$ & $38.2\pm1.5$ & $48.7\pm1.9$ & $51.9\pm1.1$ & $43.7\pm1.0$ & $\textcolor{white}{-0}57.7\pm6.8\pm8.9$ & $\textcolor{white}{-0}61.8\pm7.5\pm3.6$ \\\hline
$4.338$ & $72.9\pm1.5$ & $51.4\pm2.1$ & $38.5\pm1.5$ & $49.4\pm2.0$ & $51.7\pm1.1$ & $43.6\pm1.0$ & $\textcolor{white}{-.00}75\pm5.6\pm11\textcolor{white}{.}$ & $\textcolor{white}{-0}79.3\pm6.2\pm4.2$ \\\hline
$4.358$ & $73.3\pm1.5$ & $53.3\pm2.1$ & $39.9\pm1.6$ & $50.9\pm2.0$ & $51.7\pm1.1$ & $43.7\pm1.0$ & $\textcolor{white}{-0}91.7\pm5.6\pm4.2$ & $\textcolor{white}{-0}92.5\pm5.9\pm6.3$ \\\hline
$4.378$ & $72.8\pm1.5$ & $51.3\pm2.1$ & $38.5\pm1.5$ & $49.2\pm2.0$ & $51.6\pm1.1$ & $43.5\pm1.0$ & $\textcolor{white}{-}101.6\pm6.1\pm4.7$ & $\textcolor{white}{-0}99.7\pm6.4\pm6.8$ \\\hline
$4.387$ & $73.2\pm1.5$ & $53.3\pm2.1$ & $39.8\pm1.6$ & $50.9\pm2.0$ & $51.8\pm1.1$ & $43.6\pm1.0$ & $\textcolor{white}{-.0}101\pm18\textcolor{white}{.}\pm4.5$ & $\textcolor{white}{-.00}64\pm18\textcolor{white}{.}\pm4.5$ \\\hline
$4.397$ & $72.6\pm1.5$ & $51.1\pm2.0$ & $38.3\pm1.5$ & $48.7\pm1.9$ & $51.5\pm1.1$ & $43.3\pm1.0$ & $\textcolor{white}{-0}81.5\pm5.9\pm3.9$ & $\textcolor{white}{-0}96.6\pm6.6\pm6.6$ \\\hline
$4.416$ & $73.1\pm1.5$ & $53.2\pm2.1$ & $39.8\pm1.6$ & $50.7\pm2.0$ & $51.5\pm1.1$ & $43.3\pm1.0$ & $\textcolor{white}{-.00}55\pm18\textcolor{white}{.}\pm2.7$ & $\textcolor{white}{-.00}49\pm17\textcolor{white}{.}\pm3.4$ \\\hline
$4.416$ & $73.1\pm1.5$ & $53.2\pm2.1$ & $39.8\pm1.6$ & $50.7\pm2.0$ & $51.5\pm1.1$ & $43.3\pm1.0$ & $\textcolor{white}{-0}75.9\pm3.8\pm3.5$ & $\textcolor{white}{-0}66.6\pm3.9\pm4.6$ \\\hline
$4.437$ & $72.9\pm1.5$ & $51.6\pm2.1$ & $38.9\pm1.6$ & $49.1\pm2.0$ & $51.4\pm1.1$ & $43.2\pm0.9$ & $\textcolor{white}{-0}61.9\pm5.1\pm3.1$ & $\textcolor{white}{-0}59.9\pm5.4\pm4.2$ \\\hline
$4.467$ & $72.8\pm1.5$ & $52.5\pm2.1$ & $39.3\pm1.6$ & $49.9\pm2.0$ & $51.1\pm1.1$ & $43.2\pm0.9$ & $\textcolor{white}{-0}19.7\pm8.9\pm1.8$ & $\textcolor{white}{-0}15.9\pm8.8\pm1.7$ \\\hline
$4.527$ & $72.6\pm1.5$ & $52.5\pm2.1$ & $39.5\pm1.6$ & $49.8\pm2.0$ & $50.9\pm1.1$ & $43.0\pm0.9$ & $\textcolor{white}{-0}18.7\pm8.1\pm1.4$ & $\textcolor{white}{-0}24.8\pm8.6\pm2.0$ \\\hline
$4.575$ & $72.5\pm1.5$ & $52.8\pm2.1$ & $39.6\pm1.6$ & $49.7\pm2.0$ & $50.9\pm1.1$ & $42.8\pm0.9$ & $\textcolor{white}{-.00}11\pm13\textcolor{white}{.}\pm1.3$ & $\textcolor{white}{-.00}16\pm12\textcolor{white}{.}\pm1.5$ \\\hline
$4.600$ & $72.3\pm1.4$ & $52.9\pm2.1$ & $39.6\pm1.6$ & $49.9\pm2.0$ & $50.8\pm1.1$ & $42.6\pm0.9$ & $\textcolor{white}{-0}23.4\pm4.1\pm1.4$ & $\textcolor{white}{-0}20.1\pm3.6\pm1.7$ \\\hline
$4.612$ & $71.8\pm1.4$ & $50.7\pm2.0$ & $38.4\pm1.5$ & $47.9\pm1.9$ & $50.7\pm1.1$ & $42.7\pm0.9$ & $\textcolor{white}{-0}22.8\pm9.9\pm1.5$ & $\textcolor{white}{-0}28.4\pm9.2\pm2.2$ \\\hline
$4.628$ & $71.8\pm1.4$ & $50.3\pm2.0$ & $38.0\pm1.5$ & $47.5\pm1.9$ & $50.6\pm1.1$ & $42.5\pm0.9$ & $\textcolor{white}{-0}27.1\pm4.5\pm1.6$ & $\textcolor{white}{-0}33.6\pm4.3\pm2.5$ \\\hline
$4.641$ & $72.0\pm1.4$ & $50.6\pm2.0$ & $38.1\pm1.5$ & $47.6\pm1.9$ & $50.6\pm1.1$ & $42.6\pm0.9$ & $\textcolor{white}{-0}34.6\pm4.5\pm1.9$ & $\textcolor{white}{-0}27.3\pm3.9\pm2.1$ \\\hline
$4.661$ & $71.7\pm1.4$ & $50.5\pm2.0$ & $37.9\pm1.5$ & $47.3\pm1.9$ & $50.4\pm1.1$ & $42.6\pm0.9$ & $\textcolor{white}{-0}28.2\pm4.5\pm1.6$ & $\textcolor{white}{-0}34.2\pm4.3\pm2.5$ \\\hline
$4.682$ & $71.7\pm1.4$ & $50.5\pm2.0$ & $38.0\pm1.5$ & $47.4\pm1.9$ & $50.3\pm1.1$ & $42.3\pm0.9$ & $\textcolor{white}{-0}29.4\pm2.6\pm1.6$ & $\textcolor{white}{-0}37.6\pm2.5\pm2.7$ \\\hline
$4.699$ & $71.5\pm1.4$ & $50.5\pm2.0$ & $38.2\pm1.5$ & $47.5\pm1.9$ & $50.3\pm1.1$ & $42.4\pm0.9$ & $\textcolor{white}{-0}28.0\pm4.5\pm1.5$ & $\textcolor{white}{-0}24.9\pm3.9\pm1.9$ \\\hline
$4.740$ & $71.7\pm1.4$ & $51.9\pm2.1$ & $39.2\pm1.6$ & $48.4\pm1.9$ & $50.2\pm1.1$ & $42.4\pm0.9$ & $\textcolor{white}{-0}33.1\pm8.4\pm1.6$ & $\textcolor{white}{-0}17.6\pm6.7\pm1.5$ \\\hline
$4.750$ & $71.6\pm1.4$ & $52.3\pm2.1$ & $39.5\pm1.6$ & $48.6\pm1.9$ & $50.2\pm1.1$ & $42.1\pm0.9$ & $\textcolor{white}{-0}21.6\pm5.2\pm1.3$ & $\textcolor{white}{-0}17.8\pm4.3\pm1.5$ \\\hline
$4.781$ & $71.6\pm1.4$ & $52.2\pm2.1$ & $39.4\pm1.6$ & $48.4\pm1.9$ & $49.9\pm1.1$ & $42.2\pm0.9$ & $\textcolor{white}{-0}15.3\pm4.5\pm1.1$ & $\textcolor{white}{-0}23.6\pm3.7\pm1.8$ \\\hline
$4.843$ & $71.2\pm1.4$ & $52.0\pm2.1$ & $39.4\pm1.6$ & $48.3\pm1.9$ & $49.9\pm1.1$ & $42.0\pm0.9$ & $\textcolor{white}{-0}21.9\pm4.7\pm1.2$ & $\textcolor{white}{-0}18.3\pm3.5\pm1.5$ \\\hline
$4.918$ & $71.0\pm1.4$ & $52.0\pm2.1$ & $39.4\pm1.6$ & $48.1\pm1.9$ & $49.2\pm1.1$ & $41.7\pm0.9$ & $\textcolor{white}{-0}22.1\pm7.4\pm1.2$ & $\textcolor{white}{-0}12.3\pm5.6\pm1.2$ \\\hline
$4.951$ & $70.5\pm1.4$ & $51.4\pm2.1$ & $38.9\pm1.6$ & $47.2\pm1.9$ & $49.2\pm1.1$ & $41.4\pm0.9$ & $\textcolor{white}{-0}17.6\pm8.0\pm1.0$ & $\textcolor{white}{-0}26.5\pm6.7\pm1.9$  \\\hline\hline 

\end{tabular}
\end{center}
\label{EFF}
\end{table*}

\end{document}

%% file: authorlist_2023-05-10.tex
M.~Ablikim$^{1}$, M.~N.~Achasov$^{5,b}$, P.~Adlarson$^{74}$, X.~C.~Ai$^{80}$, R.~Aliberti$^{35}$, A.~Amoroso$^{73A,73C}$, M.~R.~An$^{39}$, Q.~An$^{70,57}$, Y.~Bai$^{56}$, O.~Bakina$^{36}$
% ...
\\
\vspace{0.2cm}
(BESIII Collaboration)\\
{\it
\vspace{0.2cm} 
$^{1}$ Institute of High Energy Physics, Beijing 100049, People's Republic of China\\
% ...
$^{5}$ Budker Institute of Nuclear Physics SB RAS (BINP), Novosibirsk 630090, Russia\\
% ...
$^{35}$ Johannes Gutenberg University of Mainz, Johann-Joachim-Becher-Weg 45, D-55099 Mainz, Germany\\
$^{36}$ Joint Institute for Nuclear Research, 141980 Dubna, Moscow region, Russia\\
% ...
$^{39}$ Liaoning Normal University, Dalian 116029, People's Republic of China\\
% ...
$^{56}$ Southeast University, Nanjing 211100, People's Republic of China\\
$^{57}$ State Key Laboratory of Particle Detection and Electronics, Beijing 100049, Hefei 230026, People's Republic of China\\
% ...
$^{70}$ University of Science and Technology of China, Hefei 230026, People's Republic of China\\
% ...
$^{73}$ University of Turin and INFN, (A)University of Turin, I-10125, Turin, Italy; (B)University of Eastern Piedmont, I-15121, Alessandria, Italy; (C)INFN, I-10125, Turin, Italy\\
$^{74}$ Uppsala University, Box 516, SE-75120 Uppsala, Sweden\\
% ...
$^{80}$ Zhengzhou University, Zhengzhou 450001, People's Republic of China\\
\vspace{0.2cm}
% ...
$^{b}$ Also at the Novosibirsk State University, Novosibirsk, 630090, Russia\\
% ...
}

%% file: JpsiX_psi3686X_BESIII_28Jan2025.bbl
\begin{thebibliography}{99}


\bibitem{PANDA}
G.~Barucca \textit{et al.} (PANDA Collaboration) 
Eur. Phys. J. A \textbf{57}, 184 (2021).

\bibitem{AMBER}
C.~Quintans  (AMBER Collaboration), 
Few-Body Syst \textbf{63}, 72 (2022).

  
\bibitem{SPD}
I.~Savin \textit{et al.},
Eur. Phys. J. A \textbf{52}, 215 (2016).

\bibitem{EIC}
A.~Accardi \textit{et al.},
Eur. Phys. J. A \textbf{52}, 268 (2016).

\bibitem{FCC}
A.~Abada \textit{et al.}, 
Eur. Phys. J. C \textbf{79}, 474 (2019).


\bibitem{ILC}
K.~Asai \textit{et al.}, 
J. High Energy Phys. \textbf{2021}, 183 (2021).

\bibitem{CEPC}
J.~B.~Guimar\~aes da Costa \textit{et al.} (CEPC Study Group), 
arXiv:1811.10545; IHEP-CEPC-DR-2018-02; IHEP-EP-2018-01; IHEP-TH-2018-01.

\bibitem{Bodwin:1994jh}
G.~T.~Bodwin, E.~Braaten and G.~Lepage,
Phys.\ Rev.\ D \textbf{51}, 1125 (1995), Phys.\ Rev.\ D \textbf{55}, 5853 (1997) (erratum).


\bibitem{Gong:2013JpsiPsi}
B.~Gong, L.-P.~Wan, J.-X.~Wang, H.-F.~Zhang,
Phys.\ Rev.\ Lett. \textbf{110}, 042002 (2013).


\bibitem{Shao:2015JpsiPsi}
H.~S.~Shao \textit{et al.}, 
J. High Energy Phys. \textbf{2015}, 103 (2015). 

\bibitem{Baranov:2019ch}
S.~P.~Baranov, A.~V.~Lipatov,
Phys.\ Rev.\ D \textbf{100}, 114021 (2019).


\bibitem{Brambilla:2022qu}
N.~Brambilla, H.~S.~Chung, A.~Vairo, X.-P.~Wang,
Phys.\ Rev.\ D \textbf{105}, L111503 (2022).


\bibitem{Butenschoen:2023psi3686}
M.~Butenschoen, B.~A.~Kniehl,
Phys.\ Rev.\ D \textbf{107}, 034003 (2023).


\bibitem{Aubert:2001pd}
B.~Aubert \textit{et al.} (BaBar Collaboration), 
Phys. Rev. Lett. \textbf{87}, 162002 (2001).


 
\bibitem{Pakhlov:2009nj}
  P.~Pakhlov {\it et al.} (Belle Collaboration), 
  Phys.\ Rev.\ D {\bf 79}, 071101 (2009).

  
\bibitem{Briere:2004ug}
  R.~A.~Briere {\it et al.} (CLEO Collaboration), 
  Phys.\ Rev.\ D {\bf 70} (2004) 072001.
  

  
\bibitem{Gong:2019rpd}
B.~Gong, Y.~D.~Wang and J.~X.~Wang,
Chin. Phys. C \textbf{43}, 083104 (2019).



\bibitem{Ablikim:2016qzw}
M.~Ablikim \textit{et al.} (BESIII Collaboration),
Phys. Rev. Lett. \textbf{118}, 092001 (2017).


 
\bibitem{Li:2014fya} 
  Y.~J.~Li \textit{et al.}, 
  Eur.\ Phys.\ J.\ C {\bf 77}, 597 (2017).
 

\bibitem{BeamEnergy2011_2014}
  M.~Ablikim {\it et al.} (BESIII Collaboration),
  Chin. Phys. C {\bf 40}, 063001 (2016).

\bibitem{BeamEnergy2017_2019}
  M.~Ablikim {\it et al.} (BESIII Collaboration),
  Chin, Phys. C {\bf 45}, 103001 (2021).

\bibitem{BeamEnergyLuminosity2020_2021}
  M.~Ablikim {\it et al.} (BESIII Collaboration), 
 Chin. Phys. C {\bf 46}, 113003 (2022). 
 

\bibitem{Luminosity2011_2017}
  M.~Ablikim {\it et al.} (BESIII Collaboration),
  Chin. Phys. C {\bf 46}, 113002 (2022).

  

\bibitem{Ablikim:2009aa}
  M.~Ablikim {\it et al.} (BESIII Collaboration),
  Nucl.\ Instrum.\ Meth.\ A {\bf 614}, 345 (2010).

\bibitem{Yu:IPAC2016-TUYA01}
   C.~H.~Yu {\it et al.},
  Proceedings of IPAC2016, Busan, Korea, 2016,
  doi:10.18429/JACoW-IPAC2016-TUYA01.
  
  
\bibitem{Ablikim:2019hff}
  M.~Ablikim {\it et al.} (BESIII Collaboration),
  Chin. Phys. C {\bf 44}, 040001 (2020);
  J.~Lu {\it et al.}, Radiat. Detect. Technol. Methods {\bf 4}, 337 (2020);
  J. W.~Zhang {\it et al.}, Radiat. Detect. Technol. Methods {\bf 6}, 289 (2022).


\bibitem{etof}
 X.~Li {\it et al.}, Radiat. Detect. Technol. Methods {\bf 1}, 13 (2017);
 Y.~X.~Guo {\it et al.}, Radiat. Detect. Technol. Methods {\bf 1}, 15 (2017);
 P.~Cao {\it et al.}, Nucl.\ Instrum.\ Meth.\ A {\bf 953}, 163053 (2020).

 
 \bibitem{geant4}
  S.~Agostinelli {\it et al.} (GEANT4 Collaboration), 
  Nucl.\ Instrum.\ Meth.\ A {\bf 506}, 250 (2003).

  \bibitem{detvis}
  K.~X.~Huang, {\it et al.},
  Nucl.\ Sci.\ Tech. {\bf 33}, 142 (2022).
 

\bibitem{ref:kkmc}
  S.~Jadach, B.~F.~L.~Ward and Z.~Was,
  Phys.\ Rev.\ D {\bf 63}, 113009 (2001);
  Comput.\ Phys.\ Commun.\  {\bf 130}, 260 (2000).

  \bibitem{ref:evtgen}
  D.~J.~Lange,
  Nucl.\ Instrum.\ Meth.\ A {\bf 462}, 152 (2001);
  R.~G.~Ping,
  Chin. Phys. C {\bf 32}, 599 (2008).

\bibitem{pdg}
  R.~L.~Workman {\it et al.} (Particle Data Group),
  Prog. Theor. Exp. Phys. {\bf 2022}, 083C01 (2022).

\bibitem{ref:lundcharm}
  J.~C.~Chen, G.~S.~Huang, X.~R.~Qi, D.~H.~Zhang and Y.~S.~Zhu,
  Phys.\ Rev.\ D {\bf 62}, 034003 (2000);
  R.~L.~Yang, R.~G.~Ping and H.~Chen,
  Chin.\ Phys.\ Lett.\  {\bf 31}, 061301 (2014).

\bibitem{photos}
  E.~Richter-Was,
  Phys.\ Lett.\ B {\bf 303}, 163 (1993).
 


\bibitem{JpsiX_3770}
M.~Ablikim \textit{et al.} (BESIII Collaboration),
Phys. Rev. Lett. \textbf{127}, 082002 (2021).

\bibitem{ISR_formula}
    E.~A.~Kuraev and V.~S.~Fadin, Sov. J. Nucl. Phys. {\bf 41}, 466 (1985).

    

\bibitem{BESIII_etaPsi}
  M.~Ablikim {\it et al.} (BESIII Collaboration),
  J. High Energ. Phys. {\bf 2021}, 177 (2021).
 


\bibitem{BESIII_Psi2pi}
  M.~Ablikim {\it et al.} (BESIII Collaboration),  
  Phys.\ Rev.\ D {\bf 104}, 052012 (2021).

\bibitem{BESIII_Jpsi2pi0}
  M.~Ablikim {\it et al.} (BESIII Collaboration),
  Phys.\ Rev.\ D {\bf 102}, 012009 (2020).


\bibitem{BESIII_JpsiEtaPi0}
  M.~Ablikim {\it et al.} (BESIII Collaboration),
  Phys.\ Rev.\ D {\bf 91}, 112005 (2015).


\bibitem{BESIII_Jpsi2K}
  M.~Ablikim {\it et al.} (BESIII Collaboration),
  Chin.\ Phys.\ C {\bf 46}, 111002 (2022).
  
 
  
\bibitem{BESIII_Jpsi2K0}
  M.~Ablikim {\it et al.} (BESIII Collaboration),
  Phys.\ Rev.\ D {\bf 107}, 092005 (2023).
 

\bibitem{BESIII_JpsiEtaPr}
  M.~Ablikim {\it et al.} (BESIII Collaboration),
  Phys.\ Rev.\ D {\bf 101}, 012008 (2020).
  
\end{thebibliography}
